\documentclass[11pt,a4paper]{article}

\usepackage{bm, amssymb, amsmath, graphicx, mathtools, float}
\usepackage{url}
\usepackage{lineno,color}
\usepackage{graphicx}
\usepackage{fancyhdr}
\usepackage{longtable,tabularx}
\usepackage{multirow}
\usepackage{subfig}
\usepackage[breaklinks,bookmarks=false]{hyperref}
\usepackage[backend=bibtex,sorting=none,maxbibnames=99,firstinits=true]{biblatex}

\fancypagestyle{firststyle}
{
	\fancyhf{}
	\fancyhead[RE,LO]{Journal of Guidance, Control, and Dynamics, Vol. 44, No. 11, pp. 1909--1923, 2021. \\
	doi: 10.2514/1.G005864}
	\fancyfoot[RE,LO]{doi: 10.2514/1.G005864}
	\fancyfoot[LE,RO]{Page \thepage}
}

\title{Approximate Analytical Solution to \\ the Zonal Harmonics Problem \\ Using Koopman Operator Theory}

\author{David Arnas\thanks{Massachusetts Institute of Technology, MA, USA. Email: \textsc{arnas@mit.edu}}, Richard Linares\thanks{Massachusetts Institute of Technology, MA, USA. Email: \textsc{linaresr@mit.edu}}}

\begin{document}

 \date{}	
 
 \maketitle{} 	
 
 \thispagestyle{firststyle}
 
\begin{abstract}
This work introduces the use of the Koopman operator theory to generate approximate analytical solutions for the zonal harmonics problem of a satellite orbiting a non-spherical celestial body. Particularly, the solution proposed directly provides the osculating evolution of the system under the effects of any order of the zonal harmonics, and can be automated to obtain any level of accuracy in the approximated solution. Moreover, this paper defines a modified set of orbital elements that can be applied to any kind of orbit and that allows the Koopman operator to have a fast convergence. In that regard, several examples of application are included, showing that the proposed methodology can be used in any kind of orbit, including circular, elliptic, parabolic and hyperbolic orbits. 
\end{abstract}


\section*{Nomenclature}

{\renewcommand\arraystretch{1.0}
	\noindent Koopman operator nomenclature
	\noindent\begin{longtable*}{@{}l @{\quad=\quad} l@{}}
		$\epsilon$  & small parameter \\
		$\phi$ & general eigenfunction \\
		$\Phi$ & matrix of eigenfunctions \\
		$\Xi$ & domain of definition of the basis functions \\
		$C_1, C_2$ & arbitrary constants \\
		$b$ & number of basis functions used \\
		$d$ & number of dimensions of the system \\
		$E$  & matrix of eigenvalues \\
		$f$ &    dynamical model \\
		$g$ & generic functional \\
		$i, j$ & matrix and vector indices \\		
		$K$  & Koopman matrix \\
		$\mathcal{K}$  & Koopman operator \\
		$L$ & set of basis functions \\
		$P_n$ & Legendre polynomial of order $n$ \\
		$t$ & time evolution [s] \\
		$T$  & Koopman modes \\
		$t_0$ & initial time [s] \\
		$V$  & matrix of eigenvectors \\
		$w$ & weighting of the basis functions \\
		${\bf x}$  & state of the system \\
		${\bf x}_0$   & initial state conditions \\
\end{longtable*}

\noindent Astrodynamics nomenclature
\noindent\begin{longtable*}{@{}l @{\quad=\quad} l@{}}
	$\alpha, p_r, s, \gamma, I_{\theta}, \beta, \xi, p_{\lambda}$ & orbital elements from Arnas and Linares~\cite{zonal} \\
	$\varLambda, \eta, \sigma, \Gamma, \kappa, \lambda, \rho$ & orbital elements for close-to-equatorial orbits \\
	$\varLambda, \eta, s, \gamma, \kappa, \beta, \chi, \rho$ & orbital elements for the general case study \\
	$\theta, \tau$ & time normalization variables \\
	$\lambda$ & longitude [rad] \\
	$\mu$ & Earth gravitational constant [$m^3/s^2$] \\
	$\nu$ & true anomaly [rad] \\
	$\varphi$ & latitude [rad] \\
	$\psi$ & normalization value for the latitude \\
	$\omega$ & argument of perigee [rad] \\
	$\Omega$ & right ascension of the ascending node [rad] \\
	$a$ & semi-major axis [$m$] \\
	$e$ & eccentricity \\
	$\mathcal{H}$ & Hamiltonian [$m^2/s^2$] \\
	$inc$ & inclination [rad] \\
	$J_n$ & zonal term of order $n$ of the Earth gravitational potential \\
	$m$ & maximum order of the zonal harmonics problem \\
	$p_{\theta}$ & modulus of the angular momentum [$m^2/s$] \\
	$p_{\lambda}$ & conjugate momenta of the longitude [$m^2/s$] \\
	$p_{\varphi}$ & conjugate momenta of the latitude [$m^2/s$] \\
	$p_r$ & conjugate momenta of the radial distance [$m/s$] \\
	$r$ & radial distance [$m$] \\
	$R_{\oplus}$ & Earth equatorial radius [$m$] \\
	$si$ & sine of the osculating inclination of the orbit \\
	$u$ & argument of latitude [rad]\\
\end{longtable*}} 
 
\newpage
\section{Introduction}

The study of the effects of the zonal harmonics on the motion of a satellite is one of the most elemental problems in celestial mechanics and is of great importance both for the design and control of space missions. However, despite being this problem one of the simplest models to study in astrodynamics (apart from the Keplerian motion), the problem is known to have no analytical solution~\cite{irigoyen1993non,celletti1995non}, and thus, a wide variety of approaches have appeared over the years to provide approximate solutions to this problem. Particularly, analytic and semi-analytic solutions are of special interest since they allow us to have a deeper understanding of the problem, to obtain a faster long term propagation of orbiting objects, to analyze the stability of orbits, or to define and assess satellite constellations and formation flying among other topics. 

From all the zonal terms of the gravitational potential, the study of the motion of a satellite subjected to the $J_2$ term of the Earth gravitational potential has always been of special importance as it is the most important gravitational perturbation for satellites orbiting the Earth. Examples of that include the solutions proposed by Brouwer~\cite{brouwer}, Kozai~\cite{kozai1959motion}, Deprit~\cite{deprit1969}, or Liu~\cite{liu1974satellite}, which are still extensively used both in theoretical and applied problems. Particularly, Brouwer~\cite{brouwer}, based on the von Zeipel perturbation method, proposed a first order solution which was later complemented by Lyddane~\cite{lyddane1963small} and Cohen and Lyddane~\cite{cohen1981radius} to study orbits with either small eccentricity or inclination, and by Coffey et al.~\cite{coffey1986critical} to be able to effectively assess orbits close to the critical inclination. Kozai~\cite{kozai1959motion} proposed, on the other hand, a different approach based on the decomposition of the motion in first-order secular, second-order secular, short-periodic, and long-periodic terms. Afterwards, he extended Brouwer's formulation by introducing a second order solution~\cite{kozai1962second} to the main satellite problem. Later, Deprit~\cite{deprit1969} approached the problem differently by making use of Lie series to define a set of canonical mappings that allows to generate an approximate solution following an iterative set of transformations~\cite{kamel1969expansion}. This methodology led to the Lie-Deprit methods, a set of perturbation techniques that has been extensively used in the literature to deal with the main satellite problem. Important examples of application of this approach include the method of elimination of the parallax~\cite{deprit1981elimination}, or the elimination of the perigee~\cite{perigee}.

In this work, we approach the problem from a different perspective. Instead of performing near unitary transformations using a small parameter, we make use of operator theory to generate solutions to this problem directly. Operator theory is based on functional analysis~\cite[Chapter~2]{conway2019course} and focuses on the study of linear operators in functional spaces~\cite[Chapter~4]{kowalski1991nonlinear}. Operator theory is widely used in theoretical physics~\cite[Chapter~7]{naylor2000} (specially in quantum mechanics~\cite[Chapter~1]{prigogine}) and other research fields~\cite{williams2015data} such as fluid-dynamics~\cite{mezic2013analysis} or control~\cite{brunton2016koopman,surana2016koopman}, but is not widely used in astrodynamics. Particularly, this paper focuses on the use of the Koopman operator, a linear operator introduced by Koopman~\cite{koopman1931hamiltonian} and later developed by Newmann~\cite{neumann1932operatorenmethode} that is able to transform a non-linear system in a finite number of dimensions into a linear system in an infinite number of dimensions, on the zonal harmonics problem around an oblate celestial body. However, due to our limitations to study a system in an infinite number of dimensions, we have to focus on a subspace of this infinite space domain, and thus, an approximation of the system is obtained instead. Nevertheless, this allows to obtain a linearization of the system (benefiting from the interesting properties of linear systems) while obtaining a good approximation to the solution. The Koopman operator has already been proposed to study problems in celestial mechanics, particularly the motion around libration points in the third body problem~\cite{libration}, and attitude dynamics~\cite{chen2020koopman}, but it has never been successfully applied to the zonal harmonics problem around a celestial body.

To that end, this work presents the methodology to apply the Koopman operator to the zonal harmonics problem. To be more precise, the contributions of this work are as follows: first, definition of a modified set of orbital elements for the specific application of the Koopman operator; second, development of a closed-form computation of the Koopman matrix using Legendre polynomials; and third, numerical error analysis of the analytical Koopman-based solution applied to the motion of a satellite orbiting a non-spherical celestial body. In particularly, this paper makes use of a modified set of orbital elements from Arnas and Linares~\cite{zonal} especially devised for their application alongside the Koopman operator. This set of orbital elements is able to represent orbits at any eccentricity, including circular, elliptic, parabolic and hyperbolic orbits. In that regard, several examples of application are included to show the performance of the methodology for different orbits. 
Additionally, this work proposes the use of Legendre polynomials to compute the Koopman matrix of the system. This provides several advantages, including lower computation complexity, easier implementation, and a better distribution of the error in the variable range defined.

The main advantage of using the Koopman operator is that it allows studying any order of the zonal harmonics problem without modifying the methodology followed. This allows automation of the computation of the solution and still obtain an approximate analytical solution to the problem. Moreover, the Koopman operator improves the accuracy of the solution by just increasing the number of basis functions used to represent the dynamical model. As mentioned before, this is done without performing any modification in the methodology, and thus, can be automatized, which means that the level of accuracy obtained is only limited by the computational power available. In addition, the Koopman operator provides the spectral behavior of the dynamical system which can be used for different applications and to study the properties of the solution~\cite{mezic2004comparison}. Finally, the technique presented generates an approximated linear dynamical system that presents good accuracy and that can be used in other applications.

This paper is organized as follows. First, we include a summary of the Koopman operator theory and the methodology followed to apply it to the zonal harmonics problem. Second, we propose two sets of orbital elements to represent the dynamical system. The notable feature of this set of orbital elements is that they allow us to represent the differential equation governing the system in completely polynomial form, while, in addition, generate expressions for the differential equation that are quasi-linear for the application of the Koopman operator methodology. This set of orbital elements are a set of variables especially devised for their use with the Koopman operator due to their particular properties. Third, we apply this methodology to a collection of four common orbits used in astrodynamics, namely, a sun-synchronous frozen orbit, a Molniya orbit, a hyperbolic orbit, and a near-equatorial orbit. This allows us to show the performance of the methodology presented in this work and its possibilities.


\section{Koopman operator theory}

We start from the classical initial value problem, represented by an autonomous system of differential equations expressed as
\begin{equation}\label{eq:autonomous}
\left\{ \begin{tabular}{l}
    $\displaystyle\frac{d}{dt}{\bf x}(t) = {\bf f}({\bf x})$ \\
    ${\bf x}(t_0) = {\bf x_0}$ 
\end{tabular}  \right.;
\end{equation}
where ${\bf x}\in\mathbb{R}^d$ is the set of $d$ variables that evolves with time $t$ and that define the state of the system. In here, ${\bf f}: \mathbb{R}^d\rightarrow \mathbb{R}^d$ represents the mathematical model of a given dynamical system, $d$ is the number of dimensions of the problem, and ${\bf x_0}$ are the initial conditions.

Let $g(\bf x)$ be an observable of the space, or in other words, $g(\bf x)$ is a function in the set of variables $\bf x$. Then, if we study the evolution of $g(\bf x)$ through the differential equation defined by Eq.~\eqref{eq:autonomous} we obtain
\begin{equation}
\frac{d}{dt}g({\bf x}) = \mathcal{K}\left(g({\bf x})\right) = \left( \nabla_{{\bf x}} g({\bf x})\right)\frac{d}{dt}{\bf x}(t) = \left( \nabla_{{\bf x}} g({\bf x})\right){\bf f}({\bf x}),
\end{equation}
where $\nabla_{{\bf x}} g = [\partial g/\partial x_1,\partial g/\partial x_2,\dots,\partial g/\partial x_d]$, and $\mathcal{K}\left(g({\bf x})\right)$ is defined as the Koopman operator applied to an observable. This means that the evolution of any observable under the effects of the dynamical system is provided by the Koopman operator:
\begin{equation}
\mathcal{K}\left(\cdot\right) \colon \left( \nabla_{{\bf x}} \cdot\right){\bf f}({\bf x}).
\end{equation}
where $\left( \nabla_{{\bf x}} \cdot\right) = [\partial (\cdot)/\partial x_1,\partial (\cdot)/\partial x_2,\dots,\partial (\cdot)/\partial x_d]$. Note that the operator is linear~\cite{koopman1931hamiltonian}, and thus,
\begin{equation}
    \mathcal{K}\left(C_1 g_1({\bf x})+C_2 g_2({\bf x})\right)=C_1\mathcal{K}\left(g_1({\bf x})\right)+C_2\mathcal{K}\left(g_2({\bf x})\right),
\end{equation}
where $C_1$ and $C_2$ are two arbitrary constants.

The main idea of the Koopman operator methodology is to find an approximated linear representation of the full dynamical system, and then use the property of linearity to obtain the exact solution to this approximation. To that end, the Koopman operator makes use of an extended configuration space of dimension $b > d$ that is able to approximate the non-linearities of the original system. In fact, if we could work with the complete infinite Hilbert space, that is, a configuration space in an infinite number of dimensions, the non-linearities of the system would be completely represented by this linear operator. However, since we are constrained by the use of a limited set of dimensions in the extended space, the linearization only provides an approximation of the original system that is more accurate the closer the system in the extended space is to a linear one.

In order to perform this linearization, this paper makes use of the Garlekin method to obtain an approximated linearized system of equations. The Garlekin method is based on the idea of performing this approximation by finding the projection of the original system into a set of orthogonal functions that define the extended space of configuration for the Koopman operator. This projection is performed using inner products between functions. Particularly, let ${f} (\bf x)$ be the function that represents the dynamical system, and let $g({\bf x})$ be a basis function. Then, the inner product between both functions is represented by
\begin{equation}
\langle  {f}, g \rangle =\int_{\Xi} {f}({\bf x})g({\bf x}) {w}({\bf x})d\bf x,
\end{equation}
where ${w} (\bf x)$ is a weighting associated with the basis functions, and $\Xi$ is the domain in which the basis functions are defined. That way, the error resulting in the approximation is orthogonal to the configuration space defined by the set of basis functions. 

Therefore, a set of orthonormal basis functions has to be defined in the extended space of dimension $b$. To that end, this work makes use of orthonormal Legendre polynomials as basis functions. The reason for that is the multiple advantages that they provide when computing the inner products. First, they are polynomials, which simplifies the computation. Second, Legendre polynomials are defined in a bounded domain ($[-1,1]$) as opposed to other basis functions, like Hermite polynomials, which are defined in all $\mathbb{R}$. Third, their weighting function is a constant, a fact that simplifies the expression of the inner product. This allows to ease the automatic computation of all the inner products required in the method.

Let $\bf L(\bf x)$ be the set of all the basis functions of the problem in the original variables written in vector form $L({\bf x})=[L_1({\bf x}),...,L_b({\bf x})]^T$, where $L_i({\bf x})$ represents the basis function $i$ from the set. Note that $L_i({\bf x})$ represents a multidimensional basis function from the set and does not require definition in a specific ordering. Particularly, and for the case of Legendre polynomials, $L_i({\bf x})$ can be expressed as
\begin{equation}
    L_i({\bf x}) = \prod_{j=1}^{d} P_{n_j}(x_j),
\end{equation}
where $P_{n_j}(x_j)$ is the Legendre polynomial of order $n_j$ applied on variable $x_j$. Therefore, the total order of the polynomial in the basis function $L_i({\bf x})$ is
\begin{equation}
    \sum_{j=1}^{d} n_j,
\end{equation} 
being the combination of the individual orders of the Legendre polynomials in each dimension (the set $\{n_1,\dots,n_d\}$) different for each basis function considered. Then, the derivative of a basis function $L_i(\bf x)$ can be obtained using the Koopman operator,
\begin{equation}
\mathcal{K}\left(L_i\right) = \frac{d}{dt}L_i = \left( \nabla_{{\bf x}} L_i\right){\bf f},
\end{equation}
where it is important to note that $\nabla_{{\bf x}}L_i(\bf x)$ can always be represented by a combination of basis functions from $\bf L(\bf x)$ due to the orthogonality and polynomial nature of the basis functions. Therefore, our objective now is to approximate this derivative with the basis functions defined in $L(\bf x)$. This is done using the inner product operation. Let $K$ be the Koopman matrix, a matrix of size $b \times b$ whose components $K_{ij}$ contain the value of the projection of the derivative of the basis function $i$ into the basis function $j$. In other words,
\begin{equation}
K_{ij} = \langle  \left( \nabla_{{\bf x}} L_i\right){\bf f}, L_j \rangle =\int_{\Xi} \left( \nabla_{{\bf x}} L_i\right){\bf f} L_j w d\bf x.
\end{equation}
For this work, these inner products are always performed between polynomials, and thus, it is possible to directly compute the integral analytically which significantly improves the computational speed of the methodology presented. Then, by following this procedure for all the combinations of basis functions, a linear system in the form:
\begin{equation} \label{eq:linear_system}
\frac{d}{dt}{\bf L} = K \bf L.
\end{equation}
is generated, which represents the best approximated linear system of the dynamics in the chosen basis functions. Note that by using higher orders of basis functions, the Koopman matrix increases in size, but it is able to better represent the non-linearities of the original system which improves the accuracy of the solution. Additionally, it is important to note that many components of the Koopman matrix are in fact zero due to the properties of the inner product which reduces the number of integrals that have to be computed. Finally, and should we want to expand the space of basis functions, it is not required to recompute the whole Koopman matrix, just the additional terms generated due to the expansion. This also means that it is possible to obtain the representation of the Koopman matrix for a given order of basis functions if we already know the Koopman matrix when using a higher order of basis functions.

As an example of generation of a Koopman matrix, let the dynamical system in study be represented by the Duffing oscillator equation:
\begin{eqnarray}
\displaystyle\frac{dx_1}{dt} & = & x_2; \nonumber \\
\displaystyle\frac{dx_2}{dt} & = & -x_1 - \epsilon x_1^3;
\end{eqnarray}
where $x_1$ and $x_2$ are the dependent variables, $\epsilon$ is a small parameter, and $t$ is the independent variable. We will assume that the problem is properly normalized in such a way that $\{x_1,x_2\} \in[-1,1]$. Imagine that we are interested in obtaining a linear representation of this system using second order Legendre basis functions. Therefore, our space will be represented by this set of basis functions: $L_1 = 1/2$, $L_2 = \sqrt{3}/2x_1$, $L_3 = \sqrt{3}/2x_2$, $L_4 = \sqrt{5}/2(3/2x_1^2-1/2)$, $L_5 = 3/2x_1x_2$, and $L_6 = \sqrt{5}/2(3/2x_2^2-1/2)$. This means that the Koopman matrix $K$ will be a $6 \times 6$ matrix defined by
\begin{eqnarray}
    K_{ij} & = & \int_{-1}^1\int_{-1}^1 \left(\displaystyle\frac{\partial L_i}{\partial x_1}, \frac{\partial L_i}{\partial x_2}\right)\left(\begin{tabular}{c} $x_2$ \\ $-x_1 - \epsilon x_1^3$\end{tabular}\right) L_j dx_1 dx_2 = \nonumber \\
    & = & \int_{-1}^1\int_{-1}^1 \left(\displaystyle\frac{\partial L_i}{\partial x_1} x_2 - \frac{\partial L_i}{\partial x_2} \left( x_1 + \epsilon x_1^3\right) \right) L_j dx_1 dx_2,
\end{eqnarray}
which describes a set of integrals that can be solved in analytical closed form since the expression is a polynomial for any combination of $i$ and $j$. Later, we will use this property for the zonal harmonics problem, being this one of the reasons why it is required to first transform the dynamical model into a polynomial system.

Once this linearization is performed, it is only required to solve the linear system represented by Eq.~\eqref{eq:linear_system}. This can be done, for instance, using the spectral theorem when the matrix $K$ is diagonalizable. In particular, let $E_i$ and $\bf v_i$ be the eigenvalues and corresponding left eigenvectors of the matrix $K$. In addition, let $V$ be the matrix containing all the left eigenvectors in its rows. Then, by pre-multiplying $V$ in Eq.~\eqref{eq:linear_system} we obtain:
\begin{equation}
V\frac{d}{dt}{\bf L} = E K {\bf L} = E V \bf L,
\end{equation}
where $E$ is a diagonal matrix containing all the eigenvalues of the system. We know that $V$ is a constant matrix since it is composed by the eigenvectors of the system, therefore,
\begin{equation}
V\frac{d}{dt}{\bf L} = \frac{d}{dt}(V{\bf L}) = E V \bf L,
\end{equation} 
where we can perform the algebraic variable transformation ${\bf \Phi} = V \bf L$ with $\phi_i = \bf v_i \bf L$, called the Koopman canonical transform, to obtain:
\begin{equation}
\frac{d}{dt}{\bf \Phi} = E\bf \Phi,
\end{equation}
and whose solution is 
\begin{equation}\label{eq:eigen_time}
{\bf \Phi}(t) = \exp(E t){\bf \Phi}(t_0).
\end{equation}
In this regard, it is important to note that the functions ${\bf \Phi} = V \bf L$ are in fact the eigenfunctions of the space defined by the dynamical system or, in other words, they are the set of functions defining the orthogonal directions of the Hilbert subspace of dimension $b$ subjected to the dynamics of the system. 

It is important to note that in general, the Koopman matrices $K$ generated using this procedure may not be diagonalizable. In those cases, we can no longer use the spectral theorem but instead, we can use other methodologies for solving linear systems such as the Jordan normal form decomposition, or the Schur decomposition, and then solve the system of differential equations sequentially.

The objective is now to recover the original variables from the system. From Eq.~\eqref{eq:eigen_time} we can obtain the evolution of the basis functions $L$,
\begin{equation}
{\bf L} (t) = V^{-1}{\bf\Phi} = V^{-1}\exp(E t) V {\bf L} (t_0).
\end{equation}
In addition, the original variables $\bf x$ can be represented using the basis functions $\bf L$ though a projection on this basis. Let $T$ be a transformation matrix of size $d \times b$ defined as
\begin{equation}
T_{ij} = \langle x_i, L_j \rangle =\int_{\Xi} x_i L_j w d\bf x,
\end{equation}
that is, the component $T_{ij}$ represents the projection of the original variable $x_i$ with $i\in\{1,\dots,d\}$ into the basis function $j$ with $j\in\{1,\dots,b\}$. The matrix $T$ is also called the Koopman modes matrix since it relates the original variables from the problem with the eigendecomposition of the dynamics, and thus, the frequencies of the motion. Therefore, the original set of variables ${\bf x}(t)$ can be expressed as
\begin{equation} \label{eq:general_solution}
{\bf x}(t) = T {\bf L}(t) = TV^{-1}\exp(E t) V {\bf L} (\bf x_0),
\end{equation}
where $\bf x_0$ are the initial conditions of the original set of variables.


\section{Zonal formulation}

In order to efficiently apply the Koopman operator to a perturbed system, we require that the system behaves like a linear system with a small perturbation. This is due to the fact that for a real application of the Koopman operator we can only use a limited number of basis functions, and thus, the closer we are to the linear system, the faster the perturbation method will converge to the solution of the problem. Therefore, we require a formulation that is able to fully represent the zonal harmonics perturbation in the system and that is close to linear for the solution to converge quickly. In that regard, Arnas and Linares~\cite{zonal} introduced a new set of orbital elements that fulfills this set of conditions. However, due to the particularities of the Koopman operator methodology and its convergence properties, some modifications to that formulation have to be performed. For that reason, we include a summary of the formulation presented in Arnas and Linares~\cite{zonal} as a reference for this paper.

The Hamiltonian in spherical coordinates of a satellite orbiting the Earth and subjected to the zonal harmonics terms of the gravitational perturbation is
\begin{equation}
\mathcal{H} = \displaystyle\frac{1}{2}\left(p_r^2 + \frac{p_{\varphi}^2}{r^2} + \frac{p_{\lambda}^2}{r^2\cos^2(\varphi)}\right) - \frac{\mu}{r}\left(1 - \sum_{n=2}^{m}J_nP_n\left(\sin(\varphi)\right)\frac{R_{\oplus}^n}{r^n}\right),
\end{equation}
where $\mu$ is the gravitational constant of the Earth, $R_{\oplus}$ is the Earth equatorial radius, $r$ is the radial distance from the satellite to the center of the Earth, $\varphi$ is the latitude, $\lambda$ is the longitude of the satellite in the inertial frame of reference for a given instant, $J_n$ are the zonal terms of the Earth gravitational potential, $P_n(\sin(\varphi))$ are the Legendre polynomials of order $n$ applied to the variable $\sin(\varphi)$, and $m$ is the maximum order of the zonal harmonics considered. In addition, the conjugate momenta of these variables are
\begin{eqnarray}
p_r & = & \dot{r}; \nonumber \\
p_{\varphi} & = & r^2\dot{\varphi}; \nonumber \\
p_{\lambda} & = & r^2\cos^2(\varphi)\dot{\lambda};
\end{eqnarray}
where we defined $\dot{x}$ as the derivative with respect to time of variable $x$. Their associated Hamilton equations are
\begin{eqnarray} \label{eq:hamiltoneqzonal}
\displaystyle\frac{dr}{dt} & = & p_r; \nonumber \\
\displaystyle\frac{dp_r}{dt} & = & -\frac{\mu}{r^2} + \frac{p_{\varphi}^2}{r^3} + \frac{p_{\lambda}^2}{r^3\cos^2(\varphi)}  + \sum_{n=2}^m (n+1)\mu J_n P_n(\sin(\varphi))\frac{R_{\oplus}^n}{r^{n+2}}; \nonumber \\
\displaystyle\frac{d\varphi}{dt} & = & \frac{p_{\varphi}}{r^2}; \nonumber \\
\displaystyle\frac{dp_{\varphi}}{dt} & = & -\frac{p_{\lambda}^2}{r^2}\frac{\sin(\varphi)}{\cos^3(\varphi)} - \sum_{n=2}^m \mu J_n \frac{\partial P_n(\sin(\varphi))}{\partial\varphi}\frac{R_{\oplus}^n}{r^{n+1}}; \nonumber \\
\displaystyle\frac{d\lambda}{dt} & = & \frac{p_{\lambda}}{r^2\cos^2(\varphi)}; \nonumber \\
\displaystyle\frac{dp_{\lambda}}{dt} & = & 0.
\end{eqnarray}
which are highly non-linear even for the unperturbed problem and thus, they are not appropriate for their direct use of the Koopman operator. To solve this, Arnas and Linares~\cite{zonal} proposed to perform a time regularization defined by
\begin{equation} \label{eq:time_regularization}
\displaystyle\frac{d\theta}{dt} = \frac{p_{\theta}}{r^2}.
\end{equation}
where
\begin{equation}
p_{\theta} =\displaystyle\sqrt{p_{\varphi}^2 + \displaystyle\frac{p_{\lambda}^2}{\cos^2(\varphi)}},
\end{equation}
is the angular momentum of the orbit, followed by a set of variable transformations defined by

\begin{eqnarray} \label{eq:originalelements}
\alpha & = & \displaystyle\frac{p_{\theta}}{r} - \frac{\mu}{p_{\theta}} = r\sqrt{\dot{\varphi}^2 + \cos^2(\varphi)\dot{\lambda}^2} -  \displaystyle\frac{\mu}{r^2\sqrt{\dot{\varphi}^2 + \cos^2(\varphi)\dot{\lambda}^2}}; \nonumber \\
p_r & = & \dot{r}; \nonumber \\
s & = & \sin(\varphi); \nonumber \\
\gamma & = & \displaystyle\frac{p_{\varphi}}{p_{\theta}}\cos(\varphi) = \displaystyle\frac{\dot{\varphi}\cos(\varphi)}{\sqrt{\dot{\varphi}^2 + \cos^2(\varphi)\dot{\lambda}^2}}; \nonumber \\
I_{\theta} & = & \displaystyle\frac{1}{\sqrt{p_{\varphi}^2 + \frac{p_{\lambda}^2}{\cos^2(\varphi)}}} = \displaystyle\frac{1}{r^2\sqrt{\dot{\varphi}^2 + \cos^2(\varphi)\dot{\lambda}^2}}; \nonumber \\
\beta & = & \lambda - \arcsin\left(\tan(\varphi)\sqrt{\displaystyle\frac{p_{\lambda}^2}{p_{\theta}^2 - p_{\lambda}^2}}\right) = \nonumber \\
& = & \lambda - \arcsin\left(\tan(\varphi)\dot{\lambda}\sqrt{\frac{\cos^4(\varphi)}{\dot{\varphi}^2 + \cos^2(\varphi)\dot{\lambda}^2 - \cos^4(\varphi)\dot{\lambda}^2}}\right); \nonumber \\
\xi & = & \displaystyle\frac{p_{\lambda}^2}{p_{\theta}^2 - p_{\lambda}^2} = \displaystyle\frac{\cos^4(\varphi)\dot{\lambda}^2}{\dot{\varphi}^2 + \cos^2(\varphi)\dot{\lambda}^2 - \cos^4(\varphi)\dot{\lambda}^2}; \nonumber \\
p_{\lambda} & = & r^2\cos^2(\varphi)\dot{\lambda}.
\end{eqnarray}
Using this expanded set of orbital elements, the system of differential equations from Eq.~\eqref{eq:j2_hamiltonian} transforms into~\cite{zonal}:

\begin{eqnarray} \label{eq:complete_poly}
\displaystyle\frac{d\alpha}{d\theta} & = & -p_r - \sum_{n=2}^{m}\mu J_nR_{\oplus}^n\displaystyle\frac{\partial P_n(s)}{\partial s}\gamma I_{\theta}^{n+1}  \left(\alpha+\mu I_{\theta}\right)^{n-1} \left(\alpha + 2\mu I_{\theta}\right); \nonumber \\
\displaystyle\frac{dp_r}{d\theta} & = & \alpha + \sum_{n=2}^m \left(n+1\right)\mu J_n R_{\oplus}^nP_n(s)I_{\theta}^{n+1}\left(\alpha + \mu I_{\theta}\right)^n; \nonumber \\
\displaystyle\frac{d s}{d\theta} & = & \gamma; \nonumber \\
\displaystyle\frac{d\gamma}{d\theta} & = & -s - \sum_{n=2}^m \mu J_n R_{\oplus}^n \frac{\partial P_n (s)}{\partial s} p_{\lambda}^2 I_{\theta}^{n+3}\left(\alpha + \mu I_{\theta}\right)^{n-1}; \nonumber \\
\displaystyle\frac{d I_{\theta}}{d\theta} & = & \sum_{n=2}^m \mu J_n R_{\oplus}^n \frac{\partial P_n (s)}{\partial s} \gamma I_{\theta}^{n+2}\left(\alpha + \mu I_{\theta}\right)^{n-1}; \nonumber \\
\displaystyle\frac{d \beta}{d\theta} & = & -\sum_{n=2}^m \mu J_n R_{\oplus}^n \frac{\partial P_n (s)}{\partial s} \frac{1}{p_{\lambda}}s \xi I_{\theta}^{n}\left(\alpha + \mu I_{\theta}\right)^{n-1}; \nonumber \\
\displaystyle\frac{d \xi}{d\theta} & = & \sum_{n=2}^m 2\mu J_n R_{\oplus}^n \frac{\partial P_n (s)}{\partial s} \frac{1}{p_{\lambda}^2}\gamma \xi^2 I_{\theta}^{n-1}\left(\alpha + \mu I_{\theta}\right)^{n-1}; \nonumber \\
\displaystyle\frac{dp_{\lambda}}{d\theta} & = & 0.
\end{eqnarray}
which is completely polynomial for the variables considered, since $p_{\lambda}$ is a constant of the motion. As mentioned before, this property will allow us to perform the direct analytical integration of the inner products following a simple and fast process.

Unfortunately, this set of equations cannot be used directly with the Koopman operator for two reasons. First, we are looking for a Koopman operator matrix that is independent on the initial conditions since this will allow us to apply the Koopman matrix obtained by the Garlekin method to any orbit. This means that we have to include $p_{\lambda}$ as a variable in the equations, making the system from Eq.~\eqref{eq:complete_poly} not linear any more under this consideration ($p_{\lambda}$ appears dividing the expression in several derivatives). Second, when using the Garlekin method we are performing a set of inner products of functions in a given range. Therefore, we require that the variables of the system are normalized in order to bound the maximum values as much as possible. In that regard, the variable $\xi$ from the previous formulation has to be transformed since it is not bounded depending on the initial conditions of the orbit.


\section{Formulation to study the zonal harmonics problem using the Koopman operator}

In this section we introduce a modified set of elements that allows to apply the Koopman operator while obtaining a good performance in the solution. To that end, the final goal is to find a Koopman matrix that allows representation of the widest range of orbits possible. In that regard, and since the variable $\beta$ is not well defined for equatorial orbits (it presents the same definition issues that the right ascension of the ascending node of an orbit), the space of solutions will be separated in two regions. The first one covers the general formulation for orbits that are far from the equatorial orbit. In particular, this region covers any orbit whose inclination is larger than 15 degrees. On the other hand, the second region covers the close-to-equatorial orbits up to inclinations of 20 degrees. This is done in order to maintain the accuracy of the Koopman operator in all the defined domain, since it is related to the region where the basis functions are defined. That way, no matter the initial conditions of the satellite, its orbit may be studied, having, in addition, an overlap between both regions in case a mission is flying near the boundary between them. Moreover it is important to note that these regions can be extended even further at a cost of a gradual loss in precision in the solution.

\subsection{General formulation}

In this paper we propose the following modified set of orbital elements for the application of the Koopman operator in the zonal harmonics problem:
\begin{eqnarray} \label{eq:spherical2zonal}
\varLambda & = & \sqrt{\displaystyle\frac{R_{\oplus}}{\mu}}\left(\frac{p_{\theta}}{r} - \frac{\mu}{p_{\theta}}\right) = \alpha\sqrt{\displaystyle\frac{R_{\oplus}}{\mu}}; \nonumber \\
\eta & = & p_r\sqrt{\frac{R_{\oplus}}{\mu}}; \nonumber \\
s & = & \sin(\varphi); \nonumber \\
\gamma & = & \displaystyle\frac{p_{\varphi}}{p_{\theta}}\cos(\varphi); \nonumber \\
\kappa & = & \sqrt{\displaystyle\frac{\mu R_{\oplus}}{p_{\varphi}^2 + \frac{p_{\lambda}^2}{\cos^2(\varphi)}}} = I_{\theta}\sqrt{\mu R_{\oplus}}; \nonumber \\
\beta & = & \lambda - \arcsin\left(\tan(\varphi)\sqrt{\displaystyle\frac{p_{\lambda}^2}{p_{\theta}^2 - p_{\lambda}^2}}\right); \nonumber \\
\chi & = & \displaystyle\frac{p_{\lambda}}{p_{\theta}}\frac{\left(\mu R_{\oplus}p_{\varphi}^2 + \frac{\mu R_{\oplus} p_{\lambda}^2}{\cos^2(\varphi)}\right)^{3/2}}{\sin^2(\varphi) + \frac{p_{\lambda}^2}{p_{\theta}^2}\cos^2(\varphi)} = p_{\lambda}I_{\theta}^4\frac{\left(\mu R_{\oplus}\right)^{3/2}}{s^2 + \gamma^2}; \nonumber \\
\rho & = & \displaystyle\frac{p_{\lambda}}{p_{\theta}} = p_{\lambda}I_{\theta}.
\end{eqnarray}
Note that in this set of equations, we have normalized several of the original orbital elements making them non-dimensional, and substituted the variable $\xi$ for the element $\chi$ which behaves much better than the original variable because its value is more constrained for a wider variety of orbits. It is also important to remark that $\chi$ is a first integral of the unperturbed problem as in the case of the elements $\kappa$, $\beta$, and $\rho$. 

When dealing with the general zonal harmonics problem around an oblate celestial body, the derivatives of this new set of orbital elements with respect to the time regularization variable ($\theta$) are:

\begin{eqnarray} \label{eq:zonal_poly}
\displaystyle\frac{d\varLambda}{d\theta} & = & -\eta - \sum_{n=2}^{m} J_n\displaystyle\frac{\partial P_n(s)}{\partial s}\gamma \kappa^{n+1}  \left(\varLambda+\kappa\right)^{n-1} \left(\varLambda+2\kappa\right); \nonumber \\
\displaystyle\frac{d\eta}{d\theta} & = & \varLambda + \sum_{n=2}^m \left(n+1\right) J_n P_n(s)\kappa^{n+1}\left(\varLambda+\kappa\right)^n; \nonumber \\
\displaystyle\frac{d s}{d\theta} & = & \gamma; \nonumber \\
\displaystyle\frac{d\gamma}{d\theta} & = & -s - \sum_{n=2}^m J_n \frac{\partial P_n (s)}{\partial s} \rho^2 \kappa^{n+1}\left(\varLambda+\kappa\right)^{n-1}; \nonumber \\
\displaystyle\frac{d \kappa}{d\theta} & = & \sum_{n=2}^m J_n \frac{\partial P_n (s)}{\partial s} \gamma \kappa^{n+2}\left(\varLambda+\kappa\right)^{n-1}; \nonumber \\
\displaystyle\frac{d \beta}{d\theta} & = & -\sum_{n=2}^m J_n \frac{\partial P_n (s)}{\partial s} s \chi \kappa^{n-2}\left(\varLambda+\kappa\right)^{n-1}; \nonumber \\
\displaystyle\frac{d \chi}{d\theta} & = & \sum_{n=2}^m 4 J_n \frac{\partial P_n (s)}{\partial s} \gamma \chi \kappa^{n+1}\left(\varLambda+\kappa\right)^{n-1} + \sum_{n=2}^m 2 J_n \frac{\partial P_n (s)}{\partial s} \rho \chi^2 \kappa^{n-2}\left(\varLambda+\kappa\right)^{n-1} \nonumber \\
& = & \sum_{n=2}^m 2 J_n \frac{\partial P_n (s)}{\partial s} \left(2\gamma\kappa^3 + \rho\chi\right) \chi \kappa^{n-2}\left(\varLambda+\kappa\right)^{n-1}; \nonumber \\
\displaystyle\frac{d \rho}{d\theta} & = & \sum_{n=2}^m J_n \frac{\partial P_n (s)}{\partial s} \gamma \rho \kappa^{n+1}\left(\varLambda+\kappa\right)^{n-1}.
\end{eqnarray}
The former system of differential equations is completely polynomial (note also that the derivatives of the Legendre polynomials $P_n$ are also polynomials in $s$), and, in addition, all the coefficients of the polynomial are either 1 (corresponding to the unperturbed problem), or a multiple of $J_n$ by a rational number (corresponding to each one of the terms from the zonal harmonics).

\subsubsection{Transformation between orbital elements}

In general, Eq.~\eqref{eq:spherical2zonal} can be used to transform spherical coordinates to the set of orbital elements proposed. However, the element $\beta$ requires a special attention in its definition due to the non uniqueness of its definition based on the arcsine function. This problem can be solved easily by defining $\beta$ as
\begin{equation}
\beta = \begin{cases} \lambda - \arcsin\left(\tan(\varphi)\sqrt{\displaystyle\frac{p_{\lambda}^2}{p_{\theta}^2 - p_{\lambda}^2}}\right) &\mbox{if } p_{\varphi} \geq 0 \\
\lambda + \arcsin\left(\tan(\varphi)\sqrt{\displaystyle\frac{p_{\lambda}^2}{p_{\theta}^2 - p_{\lambda}^2}}\right) + \pi & \mbox{if } p_{\varphi} < 0 \end{cases}.
\end{equation}
which hares the same value as the right ascension of the ascending node of the orbit.

For the inverse transformation, that is, to transform these set of orbital elements to spherical coordinates, we use the orbital element definitions provided by Eqs.~\eqref{eq:originalelements} and~\eqref{eq:spherical2zonal}, and the geometrical relations that exist between these variables. Particularly, the spherical coordinates can be obtained following this sequence of operations:
\begin{eqnarray}
p_{\theta} & = & \displaystyle\frac{\sqrt{\mu R_{\oplus}}}{\kappa}; \nonumber \\
r & = & \displaystyle\frac{p_{\theta}}{\varLambda\sqrt{\displaystyle\frac{\mu}{R_{\oplus}}} + \displaystyle\frac{\mu}{p_{\theta}}}; \nonumber \\
p_r & = & \sqrt{\displaystyle\frac{\mu}{R_{\oplus}}}\eta; \nonumber \\
\varphi & = & \arcsin (s); \nonumber \\
p_{\varphi} & = & \displaystyle\frac{\gamma}{\cos(\varphi)}p_{\theta} ; \nonumber \\
si & = & \sqrt{s^2 + \gamma^2}; \nonumber \\
u & = & \begin{cases} \arcsin\left(\displaystyle\frac{s}{si}\right) &\mbox{if } p_{\varphi} \geq 0 \\
\pi - \arcsin\left(\displaystyle\frac{s}{si}\right) & \mbox{if } p_{\varphi} < 0 \end{cases}; \nonumber \\
\lambda & = & \begin{cases} \beta \pm \arccos\left(\displaystyle\frac{\cos(u)}{\cos(\varphi)}\right) &\mbox{if } s \geq 0 \\
\beta \mp \arccos\left(\displaystyle\frac{\cos(u)}{\cos(\varphi)}\right) & \mbox{if } s < 0 \end{cases}; \nonumber \\
p_{\lambda} & = & \rho p_{\theta};
\end{eqnarray}
where $si$ is the sine of the osculating inclination of the orbit, $u$ is the argument of latitude of the orbit at a given time, and $\pm$ and $\mp$ represent the different solutions when considering prograde (upper symbol) or retrograde (lower symbol) orbits.

\subsubsection{Particular case for $J_2$}

In the cases where only the $J_2$ perturbation is considered, the Hamiltonian of the system in spherical coordinates is given by
\begin{equation}\label{eq:j2_hamiltonian}
\mathcal{H} = \displaystyle\frac{1}{2}\left(p_r^2 + \frac{p_{\varphi}^2}{r^2} + \frac{p_{\lambda}^2}{r^2\cos^2(\varphi)}\right) - \frac{\mu}{r} + \frac{1}{2}\mu R^2_{\oplus}J_2\frac{1}{r^3}\left(3\sin^2(\varphi) - 1\right),
\end{equation}
whose associated Hamilton equations after performing the time regularization, the expansion of the configuration space and the variable transformation described for the zonal problem are:
\begin{eqnarray} \label{eq:j2_poly}
\displaystyle\frac{d\varLambda}{d\theta} & = & -\eta - 3 J_2 s \gamma \kappa^{3}  \left(\varLambda+\kappa\right) \left(\varLambda+2\kappa\right); \nonumber \\
\displaystyle\frac{d\eta}{d\theta} & = & \varLambda + \frac{3}{2} J_2 \kappa^{3}\left(\varLambda+\kappa\right)^2 (3s^2 - 1); \nonumber \\
\displaystyle\frac{d s}{d\theta} & = & \gamma; \nonumber \\
\displaystyle\frac{d\gamma}{d\theta} & = & -s - 3 J_2 s \rho^2 \kappa^{3}\left(\varLambda+\kappa\right); \nonumber \\
\displaystyle\frac{d \kappa}{d\theta} & = & 3 J_2 s \gamma \kappa^{4}\left(\varLambda+\kappa\right); \nonumber \\
\displaystyle\frac{d \beta}{d\theta} & = & - 3 J_2 s^2 \chi \left(\varLambda+\kappa\right); \nonumber \\
\displaystyle\frac{d \chi}{d\theta} & = & 12 J_2 s \gamma \chi \kappa^{3}\left(\varLambda+\kappa\right) + 6 J_2 s \rho \chi^2 \left(\varLambda+\kappa\right); \nonumber \\
\displaystyle\frac{d \rho}{d\theta} & = & 3 J_2 s \gamma \rho \kappa^{3}\left(\varLambda+\kappa\right).
\end{eqnarray}
which is again completely polynomial.

\subsection{Close-to-equatorial orbits}

Orbits that are close to the equator present two difficulties when using the former formulation. First, the variable $\beta$ is not well defined for equatorial orbits in a similar way in which the right ascension of the ascending node in Keplerian variables is not well defined for equatorial orbits. Second, the value of the variable $\chi$ can increase too much for the Garlekin method to properly approximate the variable in the ranges of application. For this reason, we use a different formulation for orbits that are close-to-equatorial. 

We start with the Hamilton equations expressed in spherical coordinates provided by Eq.~\eqref{eq:hamiltoneqzonal}. From this system of differential equations we perform a time regularization, but instead of using $\theta$ as the time regularization variable as in the previous case, we use a new independent variable $\tau$ defined by
\begin{equation} \label{eq:time_regularization_eq}
\displaystyle\frac{d\tau}{dt} = \frac{p_{\theta}}{r^2\cos^2(\varphi)},
\end{equation}
and introduce it in the system of differential equations from Eq.~\eqref{eq:hamiltoneqzonal} to obtain:

\begin{eqnarray}
\displaystyle\frac{dr}{d\tau} & = & \frac{p_r}{p_{\theta}}r^2\cos^2(\varphi); \nonumber \\
\displaystyle\frac{dp_r}{d\tau} & = & -\frac{\mu}{p_{\theta}}\cos^2(\varphi) + \frac{1}{r}\frac{p_{\varphi}^2}{p_{\theta}}\cos^2(\varphi) + \frac{1}{r}\frac{p_{\lambda}^2}{p_{\theta}}  + \nonumber \\
& + & \sum_{n=2}^m (n+1)\mu J_n P_n(\sin(\varphi))\frac{R_{\oplus}^n}{r^{n}}\frac{1}{p_{\theta}}\cos^2(\varphi); \nonumber \\
\displaystyle\frac{d\varphi}{d\tau} & = & \frac{p_{\varphi}}{p_{\theta}}\cos^2(\varphi); \nonumber \\
\displaystyle\frac{dp_{\varphi}}{d\tau} & = & -\frac{p_{\lambda}^2}{p_{\theta}}\tan(\varphi)\ - \sum_{n=2}^m \mu J_n \frac{\partial P_n(\sin(\varphi))}{\partial\varphi}\frac{1}{p_{\theta}}\frac{R_{\oplus}^n}{r^{n-1}}\cos^2(\varphi); \nonumber \\
\displaystyle\frac{d\lambda}{d\tau} & = & \frac{p_{\lambda}}{p_{\theta}}; \nonumber \\
\displaystyle\frac{dp_{\lambda}}{d\tau} & = & 0.
\end{eqnarray}

Similarly to the previous section, we perform the variable transformation for $\varLambda$, $\eta$, $s$, $\gamma$, and $\rho$, and expand the configuration space of the problem with variable $\kappa$. In this regard, note that variable $\chi$ is not required to represent this region of the space. Following these transformations, the system of differential equations becomes:
\begin{eqnarray}
\displaystyle\frac{d\varLambda}{d\tau} & = & -\eta\left(1-s^2\right) - \sum_{n=2}^{m} J_n\displaystyle\frac{\partial P_n(s)}{\partial s}\gamma \kappa^{n+1}  \left(\varLambda+\kappa\right)^{n-1} \left(\varLambda+2\kappa\right)\left(1-s^2\right); \nonumber \\
\displaystyle\frac{d\eta}{d\tau} & = & \varLambda\left(1-s^2\right) + \sum_{n=2}^m \left(n+1\right) J_n P_n(s)\kappa^{n+1}\left(\varLambda+\kappa\right)^n\left(1-s^2\right); \nonumber \\
\displaystyle\frac{d s}{d\tau} & = & \gamma\left(1-s^2\right); \nonumber \\
\displaystyle\frac{d\gamma}{d\tau} & = & -s\left(1-s^2\right) - \sum_{n=2}^m J_n \frac{\partial P_n (s)}{\partial s} \rho^2 \kappa^{n+1}\left(\varLambda+\kappa\right)^{n-1}\left(1-s^2\right); \nonumber \\
\displaystyle\frac{d \kappa}{d\tau} & = & \sum_{n=2}^m J_n \frac{\partial P_n (s)}{\partial s} \gamma \kappa^{n+2}\left(\varLambda+\kappa\right)^{n-1}\left(1-s^2\right); \nonumber \\
\displaystyle\frac{d \lambda}{d\tau} & = & \rho; \nonumber \\
\displaystyle\frac{d \rho}{d\tau} & = & \sum_{n=2}^m J_n \frac{\partial P_n (s)}{\partial s} \gamma \rho \kappa^{n+1}\left(\varLambda+\kappa\right)^{n-1}\left(1-s^2\right).
\end{eqnarray}
Note that this system of differential equations is still polynomial, but contrary to what happens in the previous formulation, the non-perturbed problem is non-linear. This may seem like a limiting factor for the application of the Koopman operator, however, the non-linearities of the unperturbed problem are in fact small when compared with the linear part, making the system nearly linear. This property allows us to directly apply the Koopman operator to this system of differential equations while still obtaining good performance in the results. Unfortunately, even with these transformations, the solution for the variables $s$ and $\gamma$ provided by the Koopman operator perform poorly as we increase the inclination of the orbit. Therefore, these orbital elements must be modified to improve the performance. This is done via a normalization of these variables.

Let $\psi$ be a constant value such that $\psi > \sin(inc)$ where $inc$ is the maximum inclination of the orbits we are interested in. In our case, and since we want to cover all the space of solutions including some overlap, we define $\psi = \sin(20^{\circ})$. Note that this value can be reduced to improve performance if we are interested in orbits with lower inclinations. In particular, the closer the value of $\psi$ is to $\sin(inc)$ of the orbit, the better is the performance obtained as long as $\psi > \sin(inc)$. On the other hand, it is not convenient to increase the value of $\psi$ over $1$ since that will make the differential equations more non-linear, reducing in consequence the performance of the solution. 

Based on this constant $\psi$ we define two transformations:
\begin{eqnarray}
\sigma & = & \displaystyle\frac{s}{\psi}; \nonumber \\
\Gamma & = & \displaystyle\frac{\gamma}{\psi};
\end{eqnarray}
and introduce them into the set of differential equations to obtain
\begin{eqnarray}
\displaystyle\frac{d\varLambda}{d\tau} & = & -\eta\left(1-\psi^2\sigma^2\right) - \nonumber \\
& - & \sum_{n=2}^{m} \psi J_n\displaystyle\frac{\partial P_n(\psi\sigma)}{\partial (\psi\sigma)}\Gamma \kappa^{n+1}  \left(\varLambda+\kappa\right)^{n-1} \left(\varLambda+2\kappa\right)\left(1-\psi^2\sigma^2\right); \nonumber \\
\displaystyle\frac{d\eta}{d\tau} & = & \varLambda\left(1-\psi^2\sigma^2\right) + \sum_{n=2}^m \left(n+1\right) J_n P_n(\psi\sigma)\kappa^{n+1}\left(\varLambda+\kappa\right)^n\left(1-\psi^2\sigma^2\right); \nonumber \\
\displaystyle\frac{d \sigma}{d\tau} & = & \Gamma\left(1-\psi^2\sigma^2\right); \nonumber \\
\displaystyle\frac{d\Gamma}{d\tau} & = & -\sigma\left(1-\psi^2\sigma^2\right) - \sum_{n=2}^m \frac{1}{\psi} J_n \frac{\partial P_n(\psi\sigma)}{\partial (\psi\sigma)} \rho^2 \kappa^{n+1}\left(\varLambda+\kappa\right)^{n-1}\left(1-\psi^2\sigma^2\right); \nonumber \\
\displaystyle\frac{d \kappa}{d\tau} & = & \sum_{n=2}^m \psi J_n \frac{\partial P_n(\psi\sigma)}{\partial (\psi\sigma)} \Gamma \kappa^{n+2}\left(\varLambda+\kappa\right)^{n-1}\left(1-\psi^2\sigma^2\right); \nonumber \\
\displaystyle\frac{d \lambda}{d\tau} & = & \rho; \nonumber \\
\displaystyle\frac{d \rho}{d\tau} & = & \sum_{n=2}^m \psi J_n \frac{\partial P_n(\psi\sigma)}{\partial (\psi\sigma)} \Gamma \rho \kappa^{n+1}\left(\varLambda+\kappa\right)^{n-1}\left(1-\psi^2\sigma^2\right);
\end{eqnarray}
which can finally be applied directly to the Koopman operator formulation for any orbit whose inclination is smaller than $\arcsin(\psi)$. Note also that now the coefficients of the polynomials of the unperturbed problem that are non-linear ($\psi^2$) are small compared to the linear term. In fact, the Koopman operator treats these terms as another perturbation of the linear system, which is why it is possible to apply this non-linear formulation.

\subsubsection{Particular case for $J_2$}

As in the general case, we include the particular case of the $J_2$ term of the gravitational potential for completeness and because it is the most important term for orbiting satellites around the Earth. In that regard the system of differential equations when taking into account just the oblateness of the Earth ($J_2$ term of the Earth gravitational potential) is
\begin{eqnarray} \label{eq:j2_equatorial}
\displaystyle\frac{d\varLambda}{d\tau} & = & -\eta\left(1-\psi^2\sigma^2\right) - 3 \psi^2 J_2 \sigma \Gamma \kappa^{3}  \left(\varLambda+\kappa\right) \left(\varLambda+2\kappa\right)\left(1-\psi^2\sigma^2\right); \nonumber \\
\displaystyle\frac{d\eta}{d\tau} & = & \varLambda\left(1-\psi^2\sigma^2\right) + \frac{3}{2}  J_2 \kappa^{3}\left(\varLambda+\kappa\right)^2\left(3\psi^2\sigma^2 - 1\right)\left(1-\psi^2\sigma^2\right); \nonumber \\
\displaystyle\frac{d \sigma}{d\tau} & = & \Gamma\left(1-\psi^2\sigma^2\right); \nonumber \\
\displaystyle\frac{d\Gamma}{d\tau} & = & -\sigma\left(1-\psi^2\sigma^2\right) - 3 J_2 \sigma \rho^2 \kappa^{3}\left(\varLambda+\kappa\right)\left(1-\psi^2\sigma^2\right); \nonumber \\
\displaystyle\frac{d \kappa}{d\tau} & = & 3 \psi^2 J_2 \sigma \Gamma \kappa^{4}\left(\varLambda+\kappa\right)\left(1-\psi^2\sigma^2\right); \nonumber \\
\displaystyle\frac{d \lambda}{d\tau} & = & \rho; \nonumber \\
\displaystyle\frac{d \rho}{d\tau} & = & 3 \psi^2 J_n \sigma \Gamma \rho \kappa^{3}\left(\varLambda+\kappa\right)\left(1-\psi^2\sigma^2\right).
\end{eqnarray}

\subsection{Evolution with respect to time}

From the previous formulations, the evolution of the system is provided as a function of the time regularization variable, $\theta$ for the general case and $\tau$ for the close-to-equatorial case. Therefore, it is of interest to show the process on how to recover the evolution of the system with respect to time. 

From the time regularization presented in Eqs.~\eqref{eq:time_regularization} and~\eqref{eq:time_regularization_eq} it is possible to obtain the derivative of time with respect to the regularization variables, particularly,
\begin{eqnarray}
\displaystyle\frac{dt}{d\theta} & = & \frac{r^2}{p_{\theta}} = \sqrt{\displaystyle\frac{R_{\oplus}^3}{\mu}}\frac{1}{\kappa\left(\Lambda + \kappa\right)^2}; \nonumber \\
\displaystyle\frac{dt}{d\tau} & = & \frac{r^2\cos^2(\varphi)}{p_{\theta}} = \sqrt{\displaystyle\frac{R_{\oplus}^3}{\mu}}\frac{1 - \sigma^2\psi^2}{\kappa\left(\Lambda + \kappa\right)^2}.
\end{eqnarray}
Therefore, since all the variables in the expressions are know as a function of the regularization variables from the solution provided by the Koopman operator, it is possible to obtain the time evolution of the system through 
\begin{eqnarray}
t & = & \int\sqrt{\displaystyle\frac{R_{\oplus}^3}{\mu}}\displaystyle\frac{d\theta}{\kappa(\theta)\left(\Lambda(\theta) + \kappa(\theta)\right)^2}; \nonumber \\
t & = & \int\sqrt{\displaystyle\frac{R_{\oplus}^3}{\mu}}\displaystyle\frac{1 - \psi^2\sigma^2(\tau)}{\kappa(\tau)\left(\Lambda(\tau) + \kappa(\tau)\right)^2}d\tau;
\end{eqnarray}
for each formulation respectively. These expressions are equivalent to Kepler's equation but for the case of orbits at any eccentricity and under the effect of the zonal harmonics. It is important to note that since this computation of $t$ is based on an integral in the time regularization variables $\theta$ and $\tau$ (which have modular nature in the solution from the Koopman operator), the value of the time obtained in this expressions is defined in an orbital period basis, and thus, $t\in[-P/2,P/2]$ if the expression are integrated analytically where $P$ is the period of the orbit. This means that in general it is required to count the number of orbital periods in order to obtain the real propagation time of the orbit.


\section{Error performance of the solution}

In this section we show the result of applying the methodology described in this work to different orbits subjected to the effects of the $J_2$ perturbation. In particular, we show four cases of application: a near circular sun-synchronous orbit, a Molniya orbit, a hyperbolic orbit, and a near-equatorial orbit. This will allow us to show the error performance of the formulation for some very characteristic orbits in astrodynamics. In addition, we also include the error performance of the Koopman operator when applying basis functions of different order, showing the convergence to the solution as we increase the order of the basis functions used. In that regard, it is important to note that the set of differential equations provided by Eqs.~\eqref{eq:j2_poly} and~\eqref{eq:j2_equatorial} are based on polynomials in odd powers. This means that only the basis functions containing terms in an odd power in the polynomial contribute to the representation of the differential equation. For that reason, in the examples provided we focus on the use of basis functions with a maximum order that is odd. 

The examples provided focus on LEO orbits because the zonal harmonics perturbation is larger in this region. Therefore, a much better performance should be expected as the altitude of the orbits increases. Moreover, all of the examples included are compared with a numerical integration of the system using a Runge-Kutta scheme of order 4-5 with a relative error control of $10^{-13}$ to show the performance of the methodology for different scenarios. This numerical integration has also been compared with a Runge-Kutta-Fehlberg scheme of order 7-8 and a Runge-Kutta of order 8 to assure the accuracy of the numerical propagation.

\subsection{Sun-synchronous orbit}

\begin{figure}[!ht]
	\centering
	{\includegraphics[width = 0.48\textwidth]{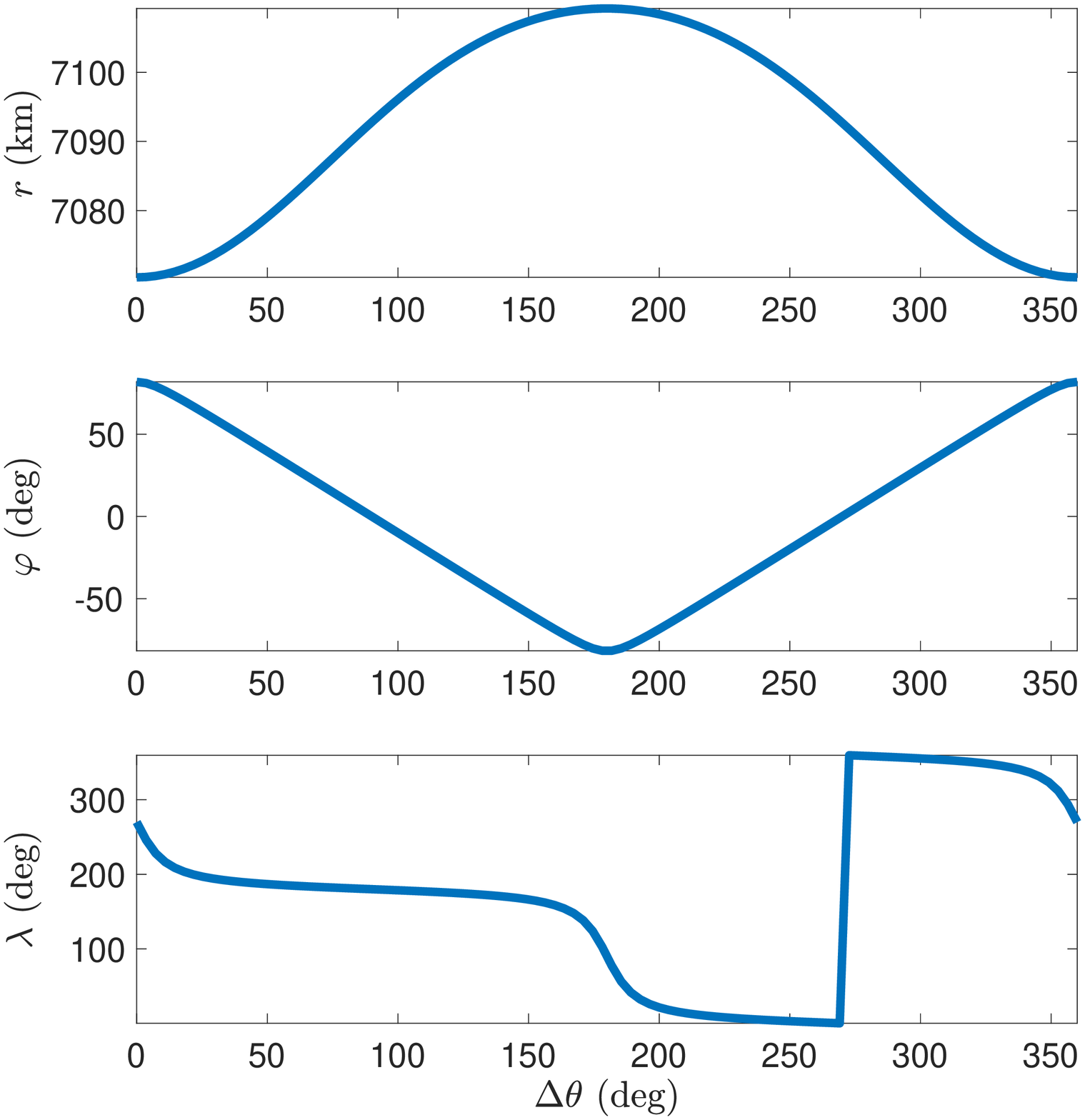}}
	\hspace{0.1cm}
	{\includegraphics[width = 0.48\textwidth]{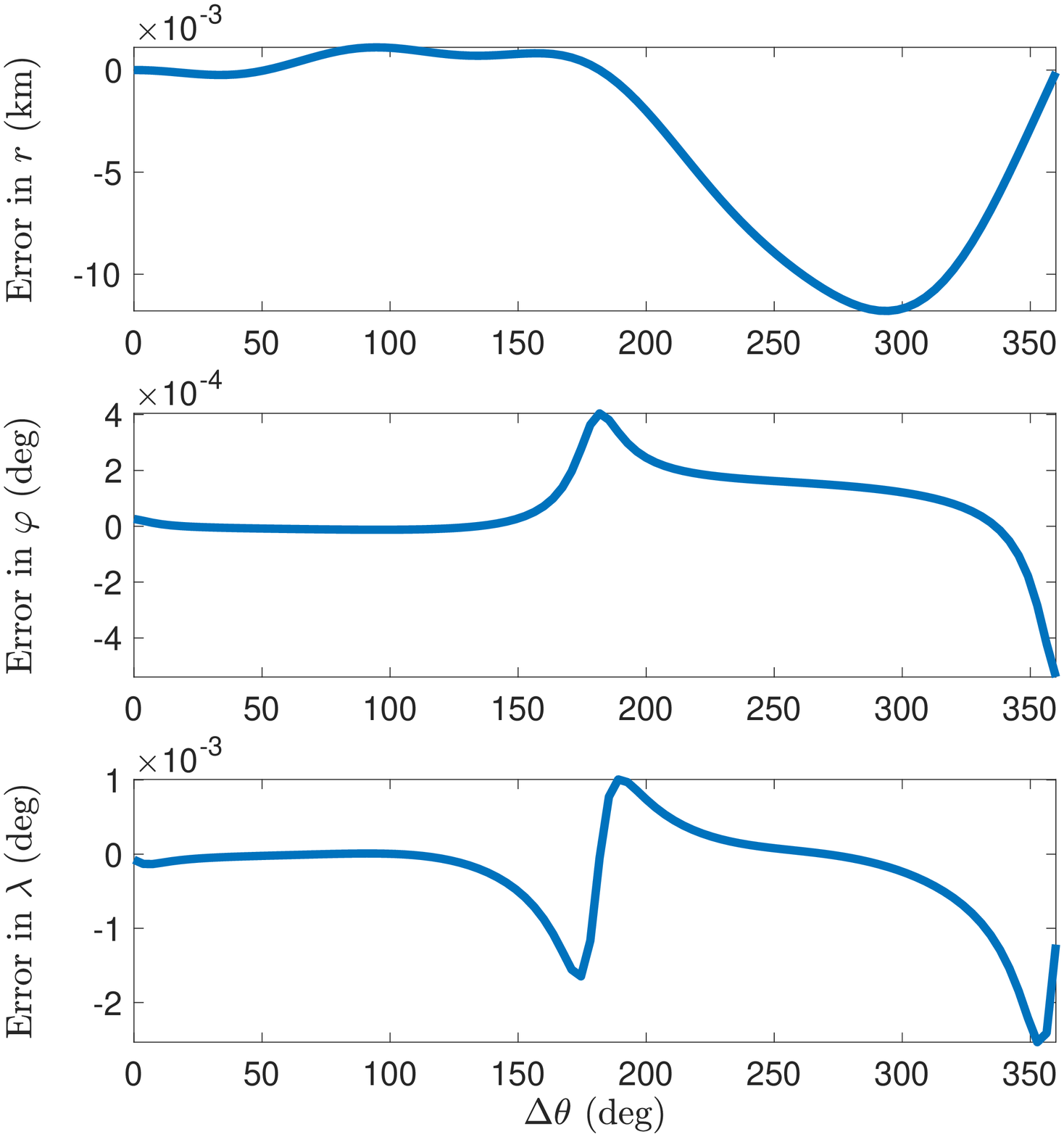}}
	\caption{7th order solution (left) and error (right) in $\{r,\varphi,\lambda\}$ for a frozen sun-synchronous orbit.}
	\label{fig:circ7}
\end{figure}

For this example, a sun-synchronous frozen Earth observation orbit is selected with the initial orbital conditions: semi-major axis $a = 7077.722$ km, eccentricity $e = 0.001043$, inclination $inc = 98.186 \deg$, argument of perigee $\omega = 90.0 \deg$, right ascension of the ascending node $\Omega = 0.0 \deg$ and true anomaly $\nu = 0.0 \deg$. Figure~\ref{fig:circ7} (left) shows the evolution of the radial distance ($r$), the latitude ($\varphi$), and the longitude ($\lambda$) of the orbit as a function of the independent variable $\theta$ (defined as the time regularization variable in Eq.~\eqref{eq:time_regularization}). Moreover, Figure~\ref{fig:circ7} (right) shows the comparison of the Koopman operator solution using basis functions of order 7 with the numerical integration of the equations. From these results, it can be observed that the error of this solution is in the order of magnitude of meters in the position of the satellite (both taking into account both the radial distance and the latitude and longitude angles of the orbit).

\begin{figure}[!ht]
	\centering
	{\includegraphics[width = 0.48\textwidth]{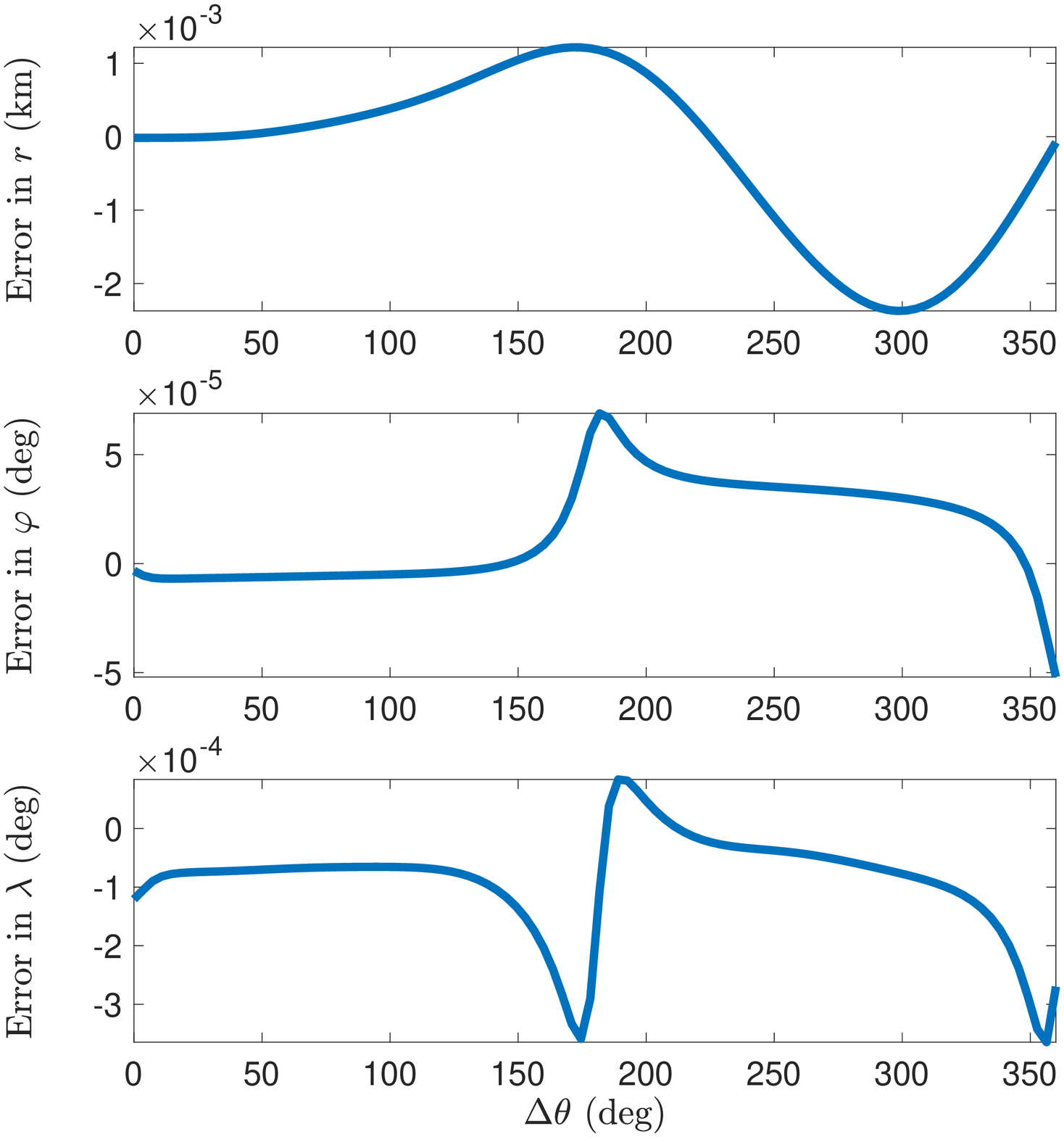}}
	\hspace{0.1cm}
	{\includegraphics[width = 0.48\textwidth]{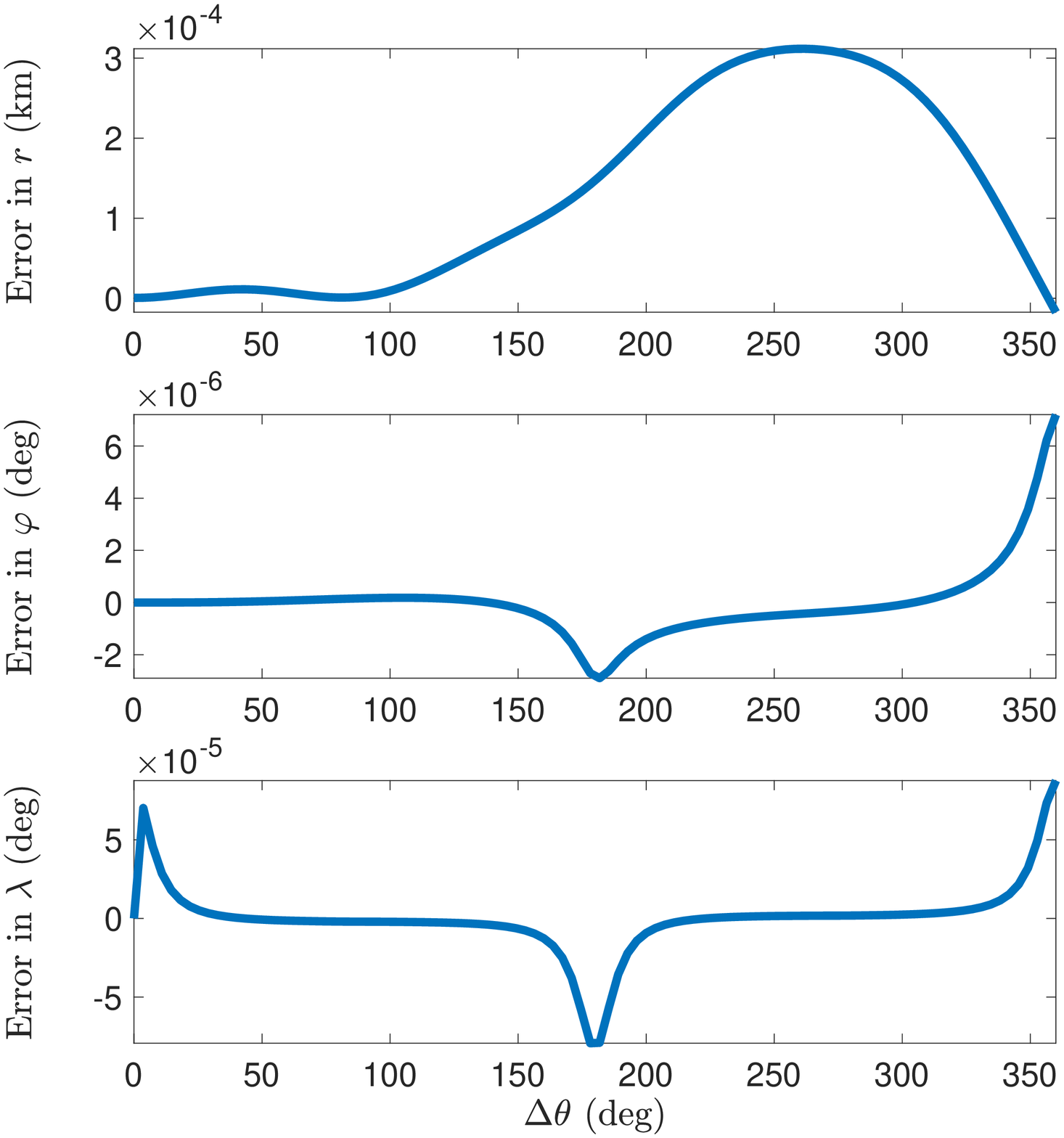}}
	\caption{9th (left) and 11th (right) order solution error in $\{r,\varphi,\lambda\}$ for a frozen sun-synchronous orbit.}
	\label{fig:circ911}
\end{figure}

The result from Figure~\ref{fig:circ7} can be improved by increasing the order of the basis functions used in the Koopman operator approach. In that regard, Figure~\ref{fig:circ911} shows the errors of the solution for the 9th (left) and 11th (right) order basis function solutions when compared to the numerical integration. As can be seen, the maximum error is reduced to 2.37 meters and less than 0.32 meters respectively. This shows that the solution is getting closer to the real evolution of the system by increasing the order of the basis functions used to represent the system of differential equations.

\begin{figure}[!ht]
	\centering
	{\includegraphics[width = 0.48\textwidth]{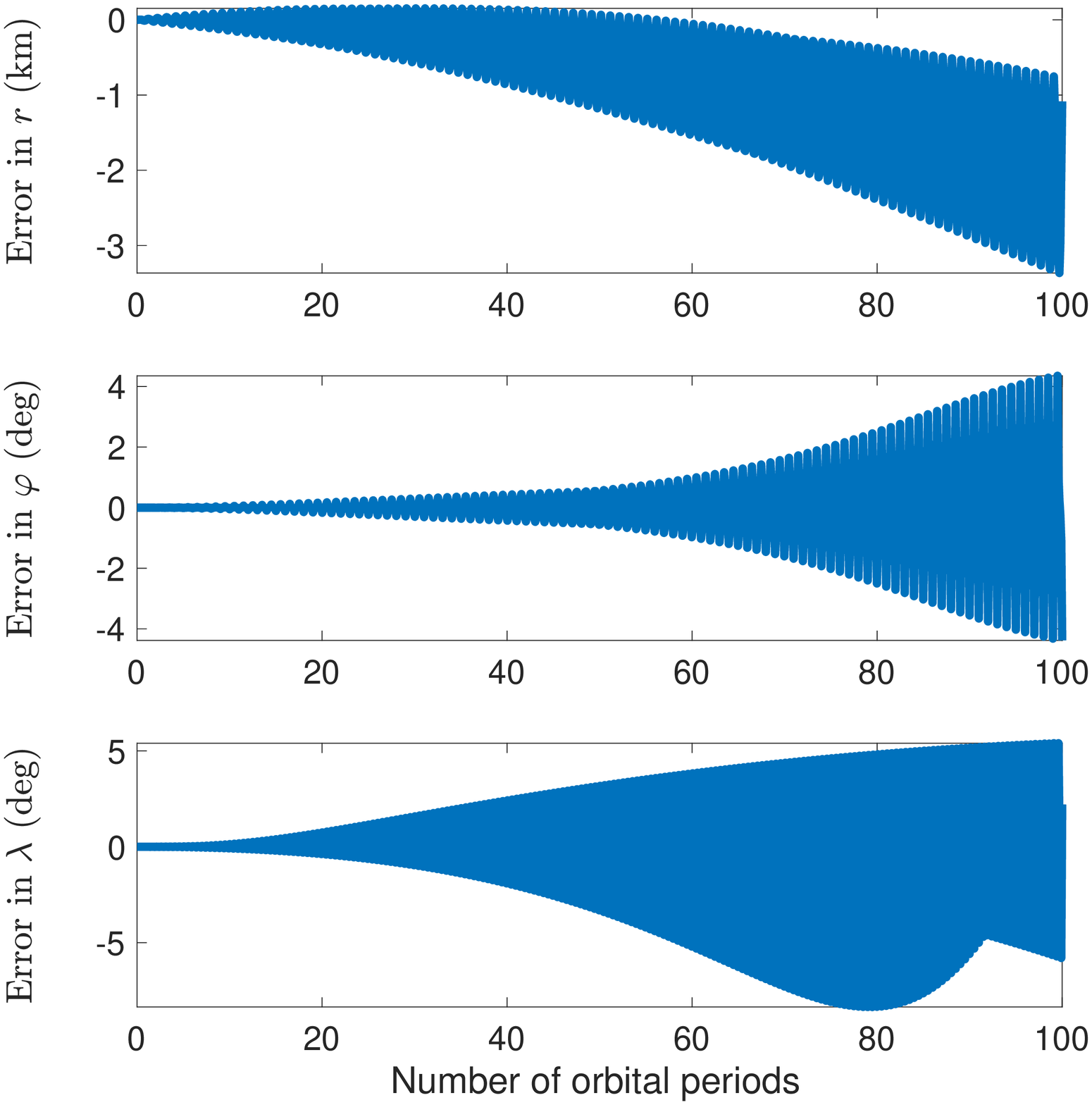}}
	\hspace{0.1cm}
	{\includegraphics[width = 0.48\textwidth]{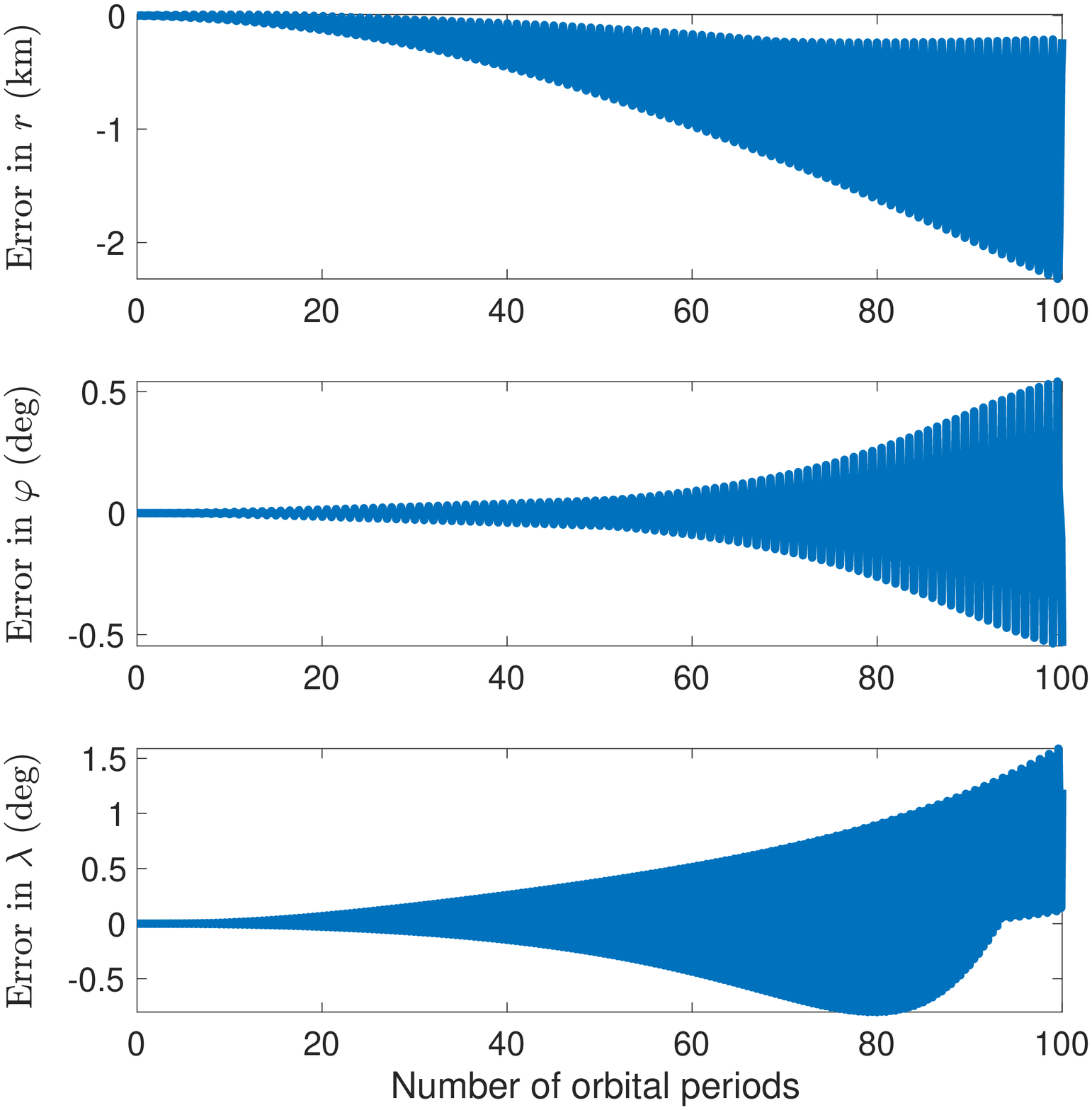}}
	\caption{7th (left) and 9th (right) order solution error for a long term propagation in $\{r,\varphi,\lambda\}$ for a frozen sun-synchronous orbit.}
	\label{fig:circlong}
\end{figure}

Moreover, we can also study the long term evolution of these solutions. To that end, Figure~\ref{fig:circlong} shows the error for a propagation of 100 orbital revolutions when using a set of basis functions of order 7th (left) and 9th (right). As it can be observed, the error of the solution quickly increases with time. Nevertheless, the error reduces with the increase on the order of the basis functions. Thus, it is possible to improve the long term performance of the methodology by further increasing the order of the basis functions used. 

\begin{figure}[!ht]
	\centering
	{\includegraphics[width = 0.48\textwidth]{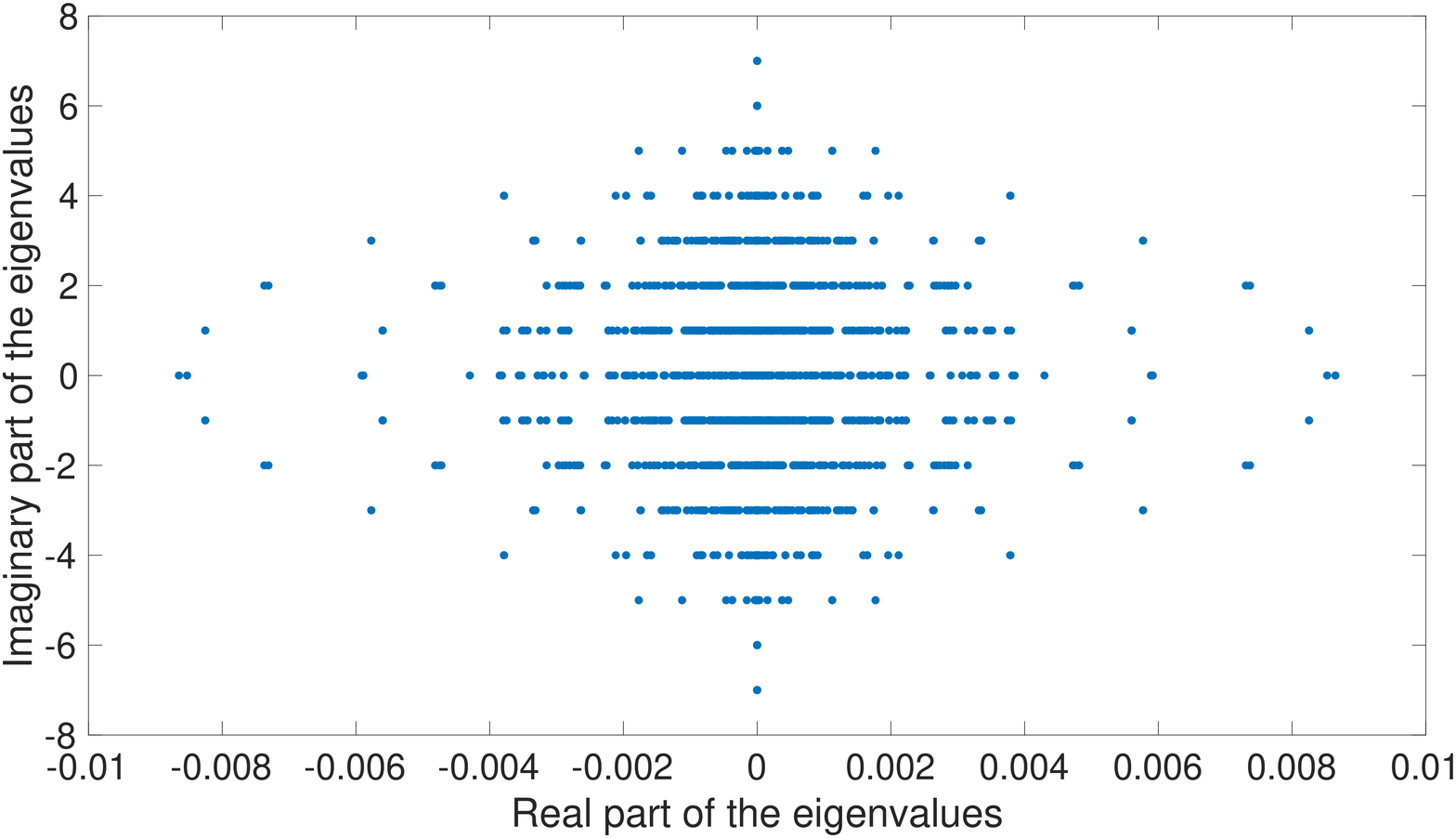}}
	\hspace{0.1cm}
	{\includegraphics[width = 0.48\textwidth]{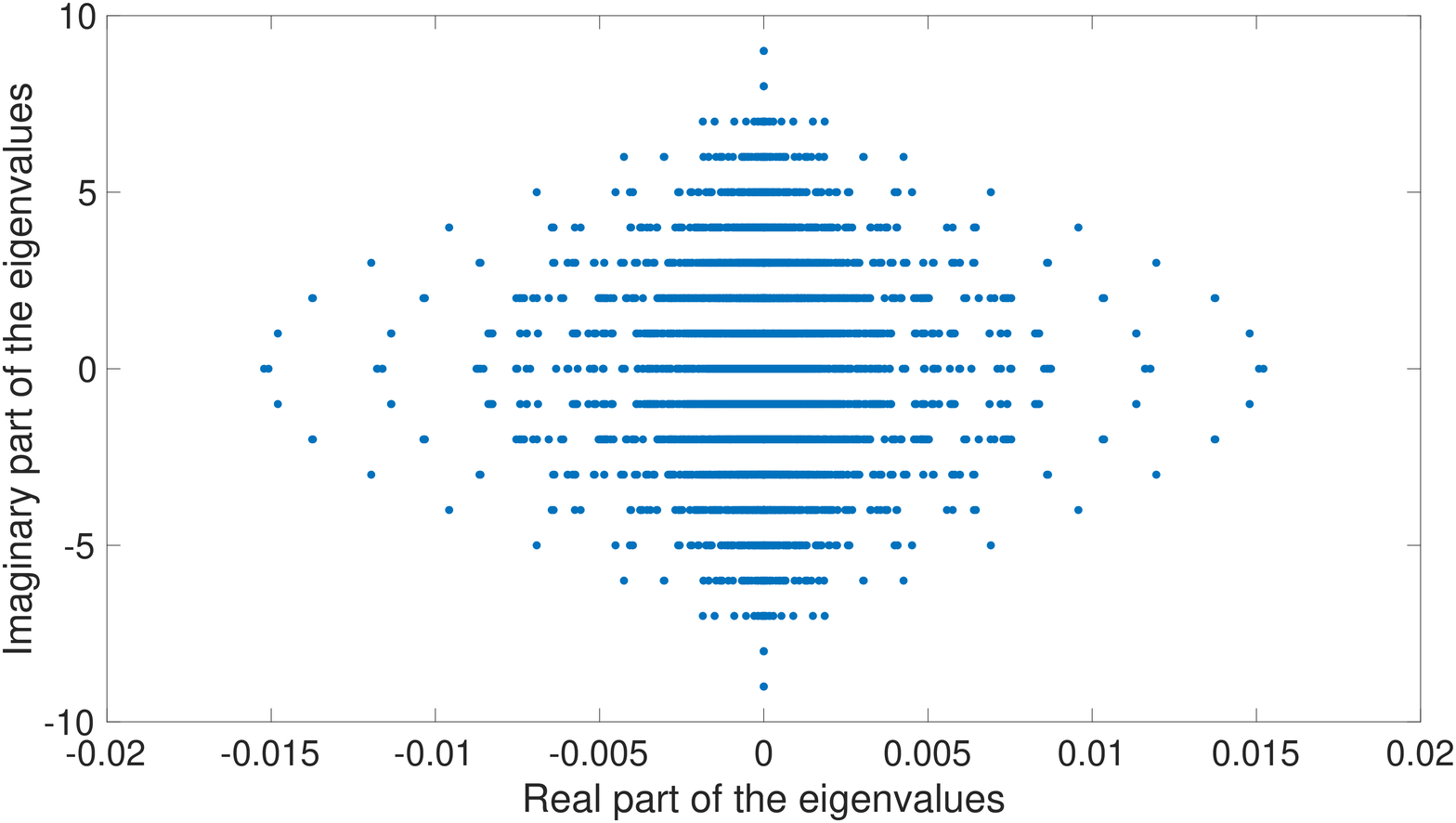}}
	\caption{7th (left) and 9th (right) order eigenvalues for the general formulation case.}
	\label{fig:eigen}
\end{figure} 

Additionally, it is also possible to study the spectral structure of the solution. To that end, Figure~\ref{fig:eigen} shows the distribution of eigenvalues of the system when using a set of basis functions of order 7 (left) and 9 (right) respectively. In that regard, it is important to note that, since the Koopman matrix is independent of the initial conditions of the orbit (matrix $K$ only depends on the dynamical model), these eigenvalues are valid for any orbit that the dynamical system provided by Eq.~\eqref{eq:j2_poly} is able to represent. As can be seen in the figure, the maximum values of the frequencies are 7 and 9 respectively. In that respect, we know that the eigenvalues of the unperturbed problem using the system from Eq.~\eqref{eq:j2_poly} are $\pm \imath$ and $0$. However, since the basis functions are built as a combination of the initial set of variables, the maximum frequency of the system increases with the order of the basis functions used. Particularly, let $\phi_1$ and $\phi_2$ be two eigenfunctions of the Koopman matrix with associated eigenvalues $E_1$ and $E_2$ respectively, therefore:
\begin{eqnarray}
\displaystyle\frac{d\phi_1}{dt} = E_1\phi_1; \nonumber \\
\displaystyle\frac{d\phi_2}{dt} = E_2\phi_2.
\end{eqnarray}
Then, we can define a third function $g$ as the multiplication of the first two, that is: $g = \phi_1 \phi_2$. If we introduce it in the dynamical system we obtain:
\begin{equation}
\displaystyle\frac{d g}{dt} = \frac{d\phi_1}{dt}\phi_2 + \phi_1\frac{d\phi_2}{dt} = E_1\phi_1\phi_2 + E_2\phi_1\phi_2 = (E_1+E_2)\phi_1\phi_2 = (E_1+E_2)g,
\end{equation}
which means that $g$ is also an eigenfunction of the system with associated eigenvalue $E_1+E_2$. This is the reason why the maximum frequency that we observe in Figure~\ref{fig:eigen} matches the maximum order of the basis functions used. Note also that the eigenvalues from both solutions are not exactly the same for two reasons. First, increasing the order of basis functions has the effect of increasing the number of eigenvalues of the system and thus, Figure~\ref{fig:eigen} (right) has more eigenvalues than Figure~\ref{fig:eigen} (left). Second, increasing the order of basis functions increases the accuracy of these eigenvalues. Therefore, small variations between the eigenvalues from both solutions are expected. On the other hand, Figure~\ref{fig:eigen} also shows that there are some eigenvalues with real part different to zero. These eigenvalues are generated by the multiple combinations of the basis functions related with the secular variation of the right ascension of the ascending node (represented by variable $\beta$) with the rest of variables involved in the problem. As it will be seen in the following example, these eigenvalues may deteriorate the solution of the method in the long term.

\subsection{Molniya orbit}

\begin{figure}[!ht]
	\centering
	{\includegraphics[width = 0.48\textwidth]{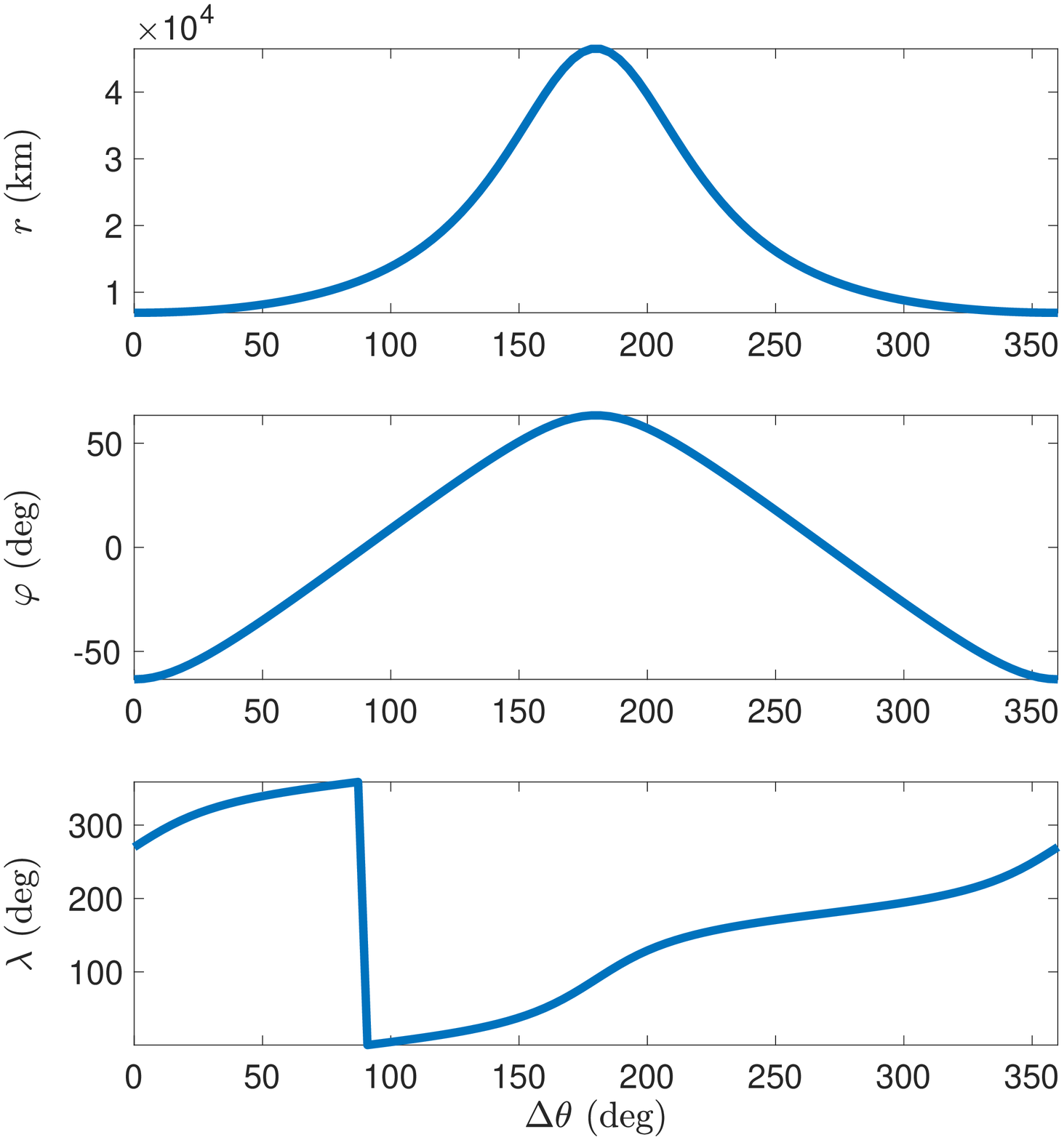}}
	\hspace{0.1cm}
	{\includegraphics[width = 0.48\textwidth]{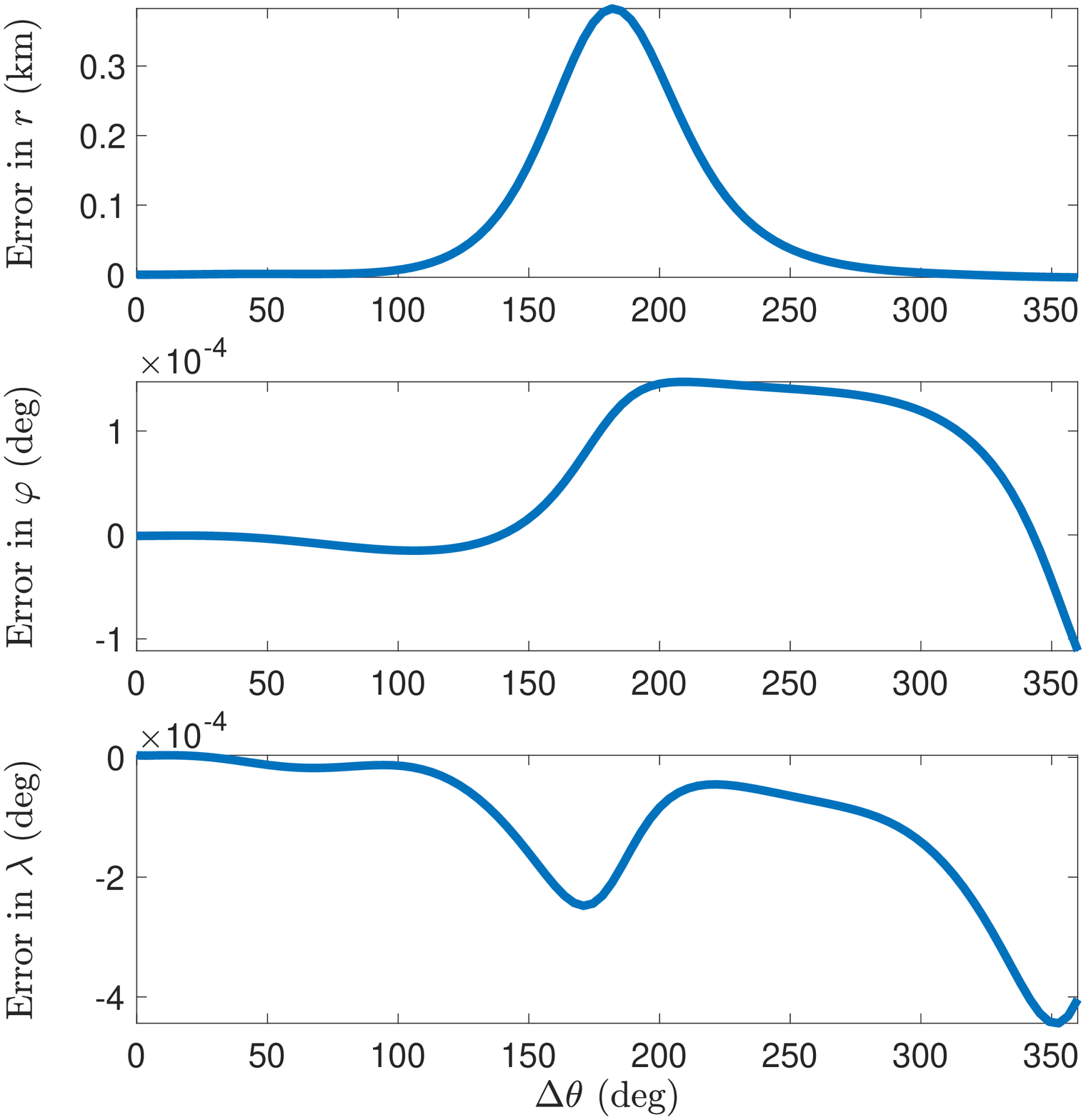}}
	\caption{7th order solution (left) and error (right) in $\{r,\varphi,\lambda\}$ for a Molniya orbit.}
	\label{fig:mol7}
\end{figure}

We select a typical Molniya orbit as an example of application to eccentric orbits. Particularly, we set the initial conditions for the propagation to: $a = 26600.0$ km, $e = 0.74$, $inc = 63.435 \deg$, $\omega = 270.0 \deg$, $\Omega = 0.0 \deg$ and $\nu = 0.0 \deg$. Figure~\ref{fig:mol7} shows the results for the evolution of $\{r,\varphi,\lambda\}$ and their error when using a set of basis functions of order 7. The maximum error observed (less than 400 m) is now bigger than in the case of the sun-synchronous orbit. Nevertheless, this is consistent with the results from the previous example having a relative error in the order of magnitude of $10^{-5}$.

\begin{figure}[!ht]
	\centering
	{\includegraphics[width = 0.48\textwidth]{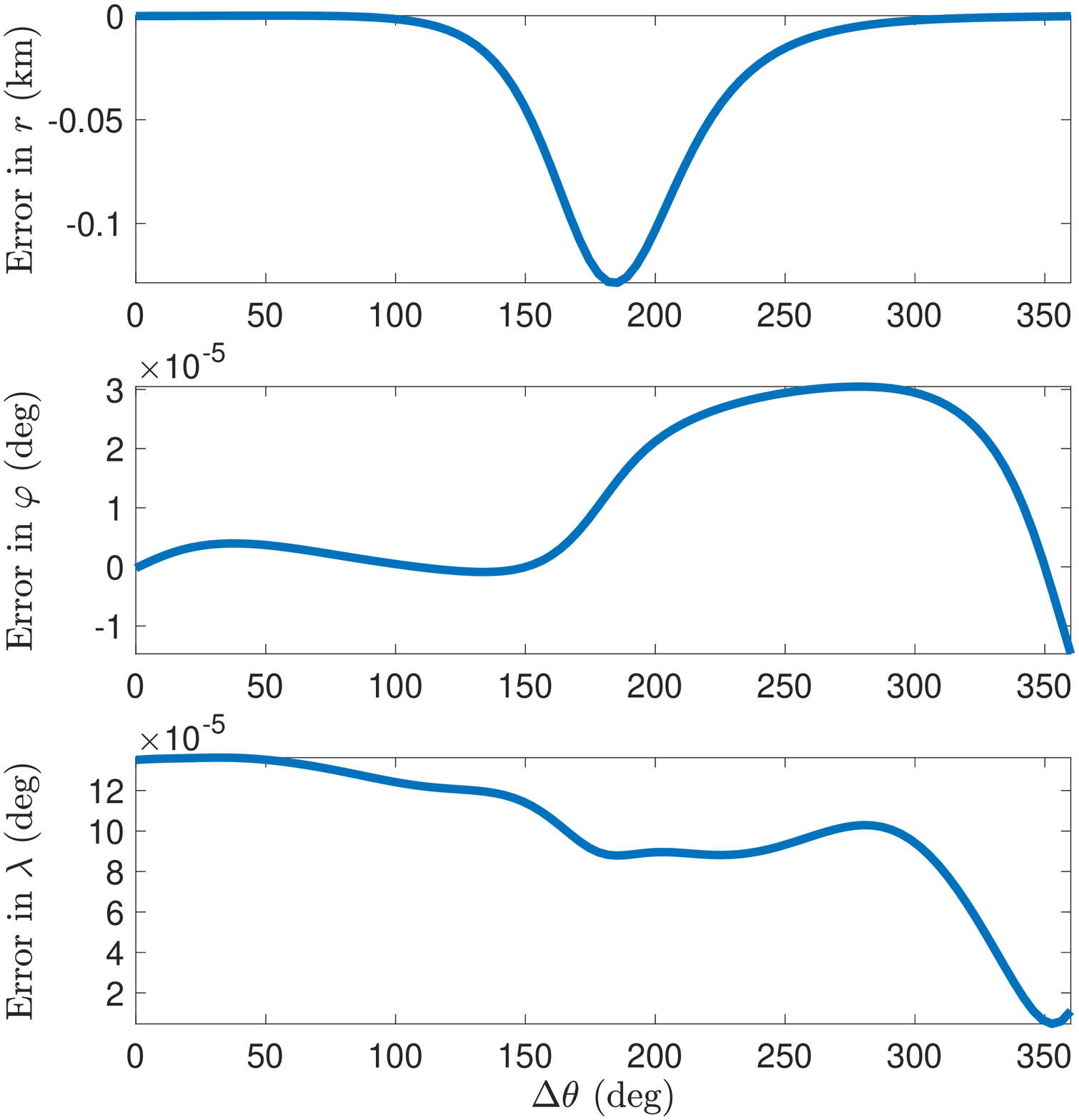}}
	\hspace{0.1cm}
	{\includegraphics[width = 0.48\textwidth]{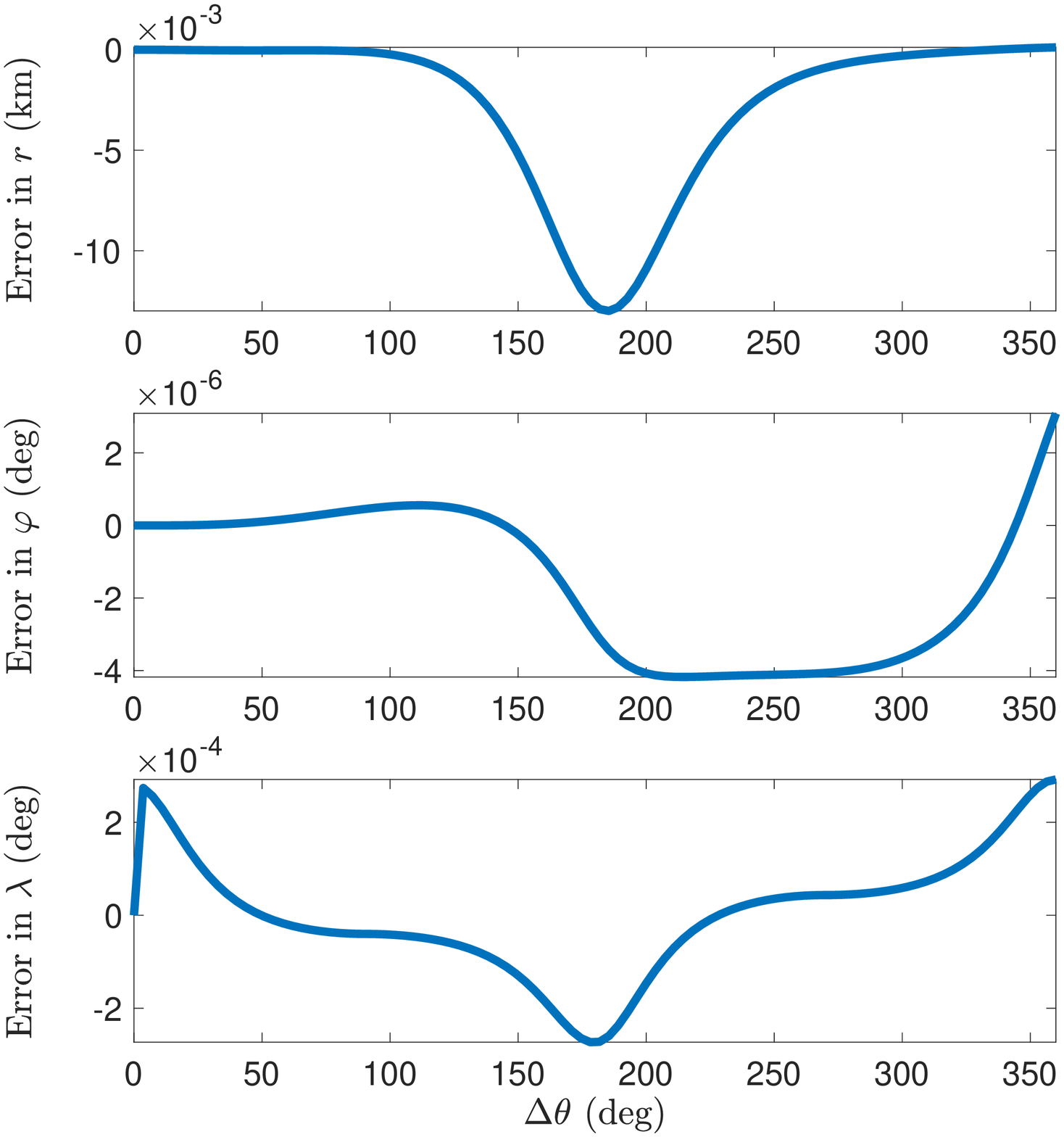}}
	\caption{9th (left) and 11th (right) order solution error in $\{r,\varphi,\lambda\}$ for a Molniya orbit.}
	\label{fig:mol911}
\end{figure} 

On the other hand, Figure~\ref{fig:mol911} shows the error on the position when using a set of basis functions of order 9 (left) and 11 (right) respectively. The maximum error in this case is 13 meters. This shows that the formulation and methodology presented can be applied to very eccentric orbits while obtaining a good accuracy.

\newpage

\begin{figure}[!ht]
	\centering
	{\includegraphics[width = 0.48\textwidth]{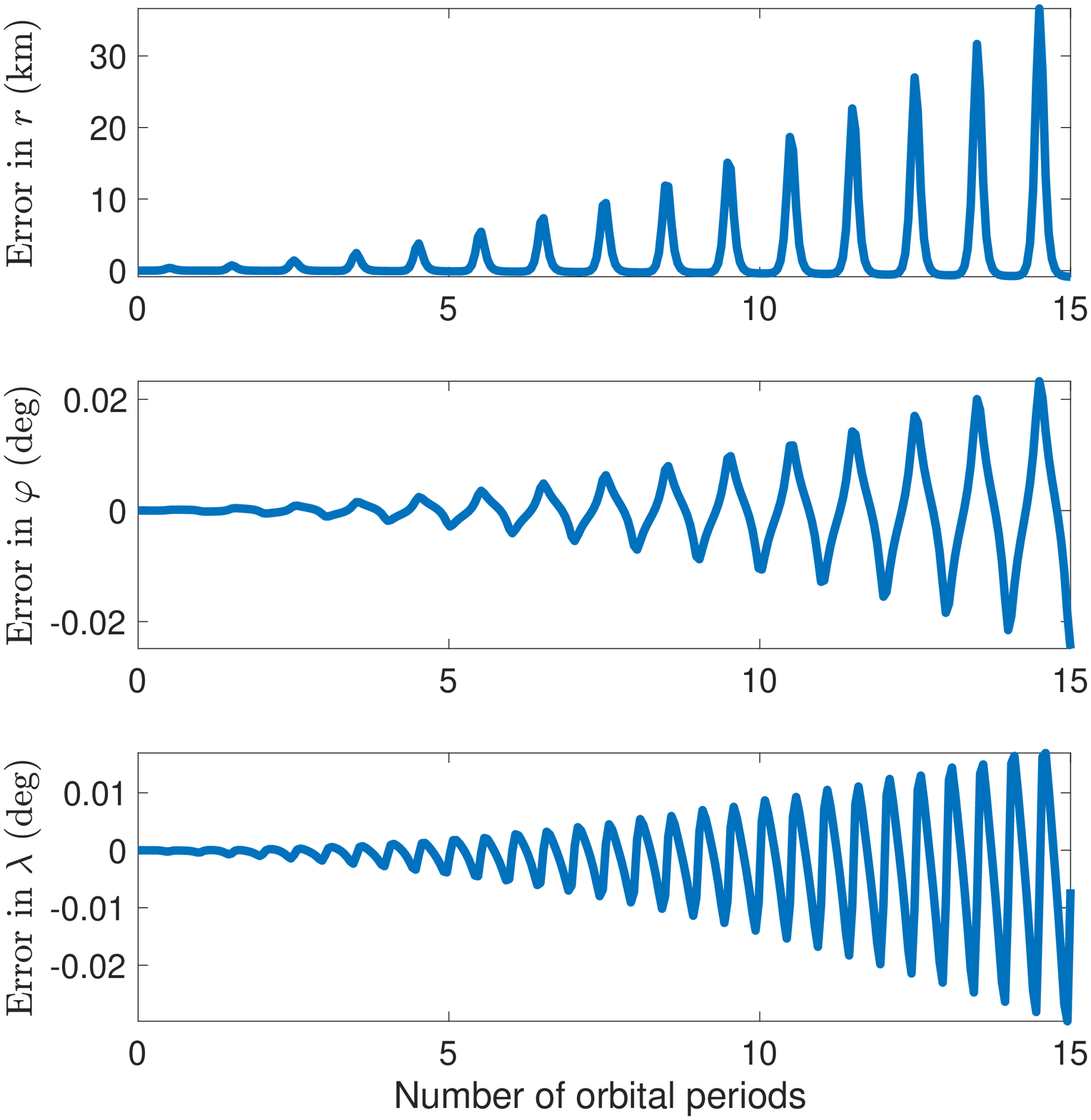}}
	\hspace{0.1cm}
	{\includegraphics[width = 0.48\textwidth]{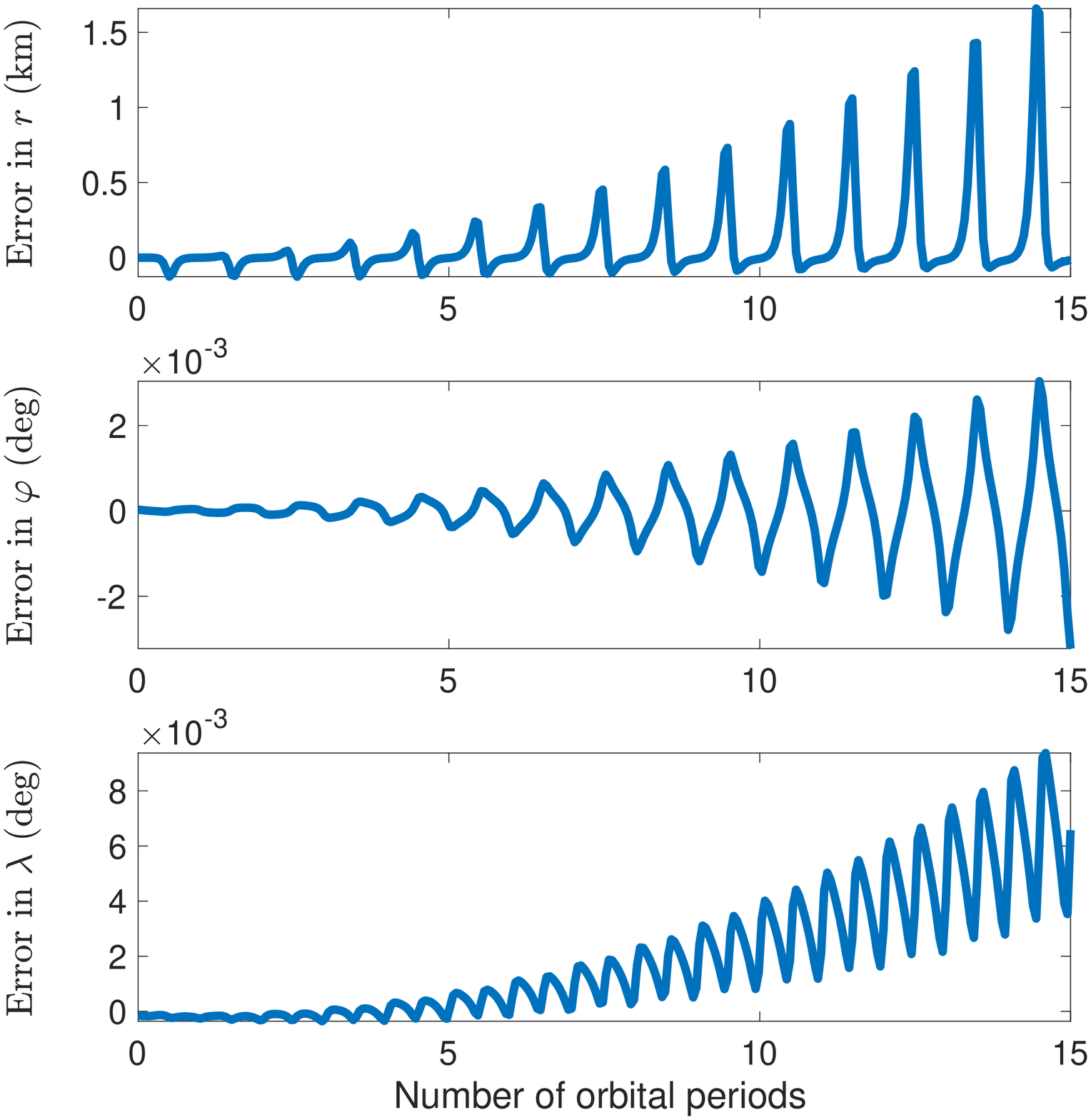}}
	\caption{7th (left) and 9th (right) order solution error for a long term propagation in $\{r,\varphi,\lambda\}$ for the Molniya orbit.}
	\label{fig:mollong}
\end{figure} 

In addition, Figure~\ref{fig:mollong} presents the long term error for this Molniya orbit for 15 orbital revolutions using basis functions of order 7 (left) and 9 (right) in the Koopman operator solution. Figure~\ref{fig:mollong} (left) shows that the error using 7th order basis functions increases very rapidly over time, reaching $37$ km at the end of the propagation, which shows the degradation of the performance of this solution in the long term propagation. On the other hand, Figure~\ref{fig:mollong} (right) shows the error using 9th order basis functions, where it can be observed that the radial error has significantly improved ($1.6$ km of maximum radial error). One important characteristic to note is that the error in the longitude of the orbit presents an exponential behavior. This effect is produced by the numerical instability that eigenvalues with positive real part introduce in the solution in the long term. In general, different initial conditions activate and distribute weighting differently between the eigenvalues of the solution through the Koopman modes $T$. Therefore, activating eigenvalues with positive real part may trigger this behavior in the solution. Nevertheless, increasing even more the order of basis functions helps to mitigate this issue but requires longer computation times and more memory. In comparison, the solution proposed in Arnas and Linares~\cite{zonal} using the Poincar\'e-Lindstedt method generates solutions with better accuracy when longer propagation times are considered. However, it is important to note that the Poincar\'e-Lindstedt method actively removes the non-desired secular effects from the solution by modifying the frequency of the variable oscillation, which, using the proposed method, is not possible since we are fixing the Koopman matrix. Future research should be made to develop alternative methods to control this behavior using the Koopman operator.

\subsection{Hyperbolic orbit}

\begin{figure}[!ht]
	\centering
	{\includegraphics[width = 0.48\textwidth]{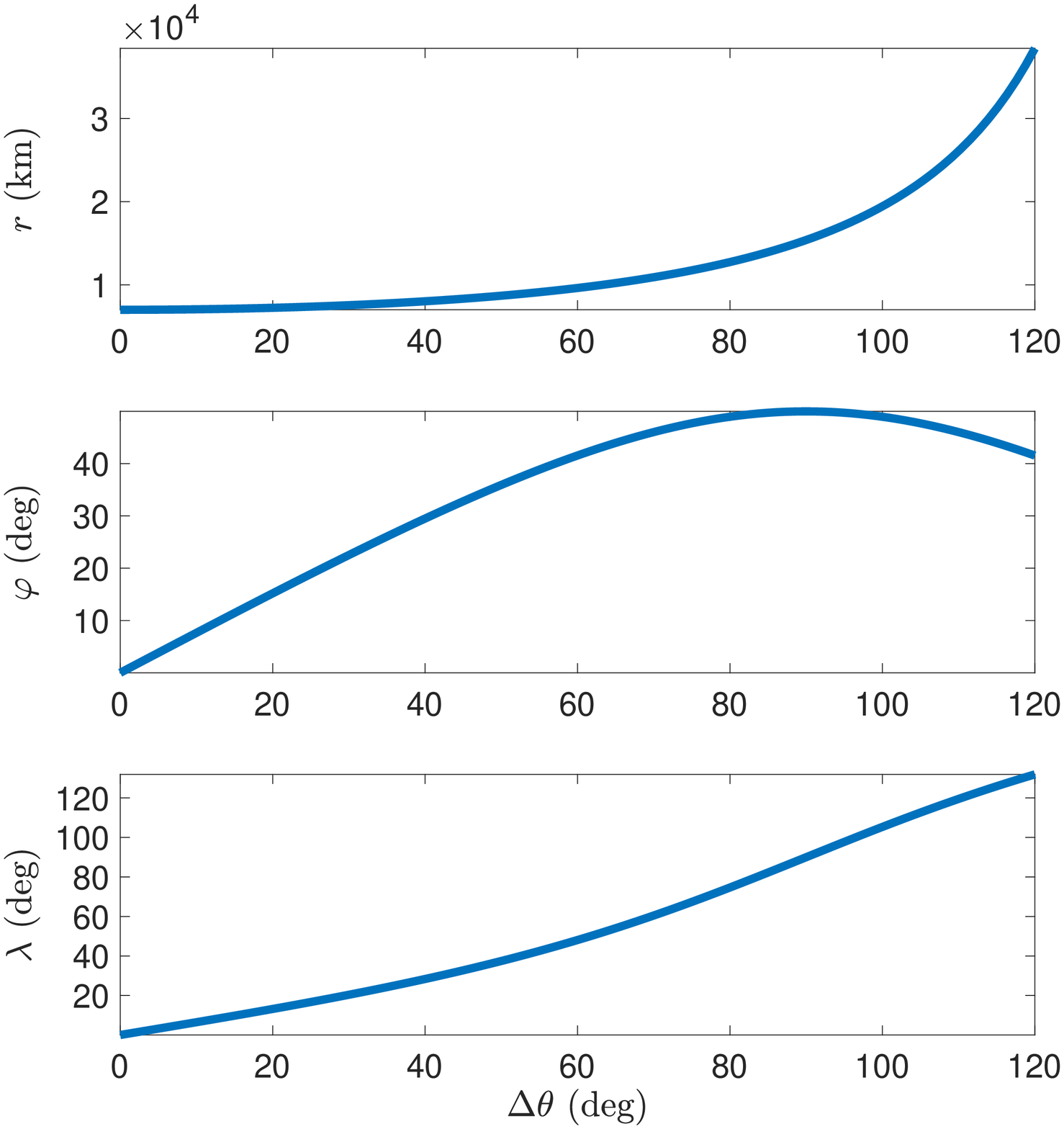}}
	\hspace{0.1cm}
	{\includegraphics[width = 0.48\textwidth]{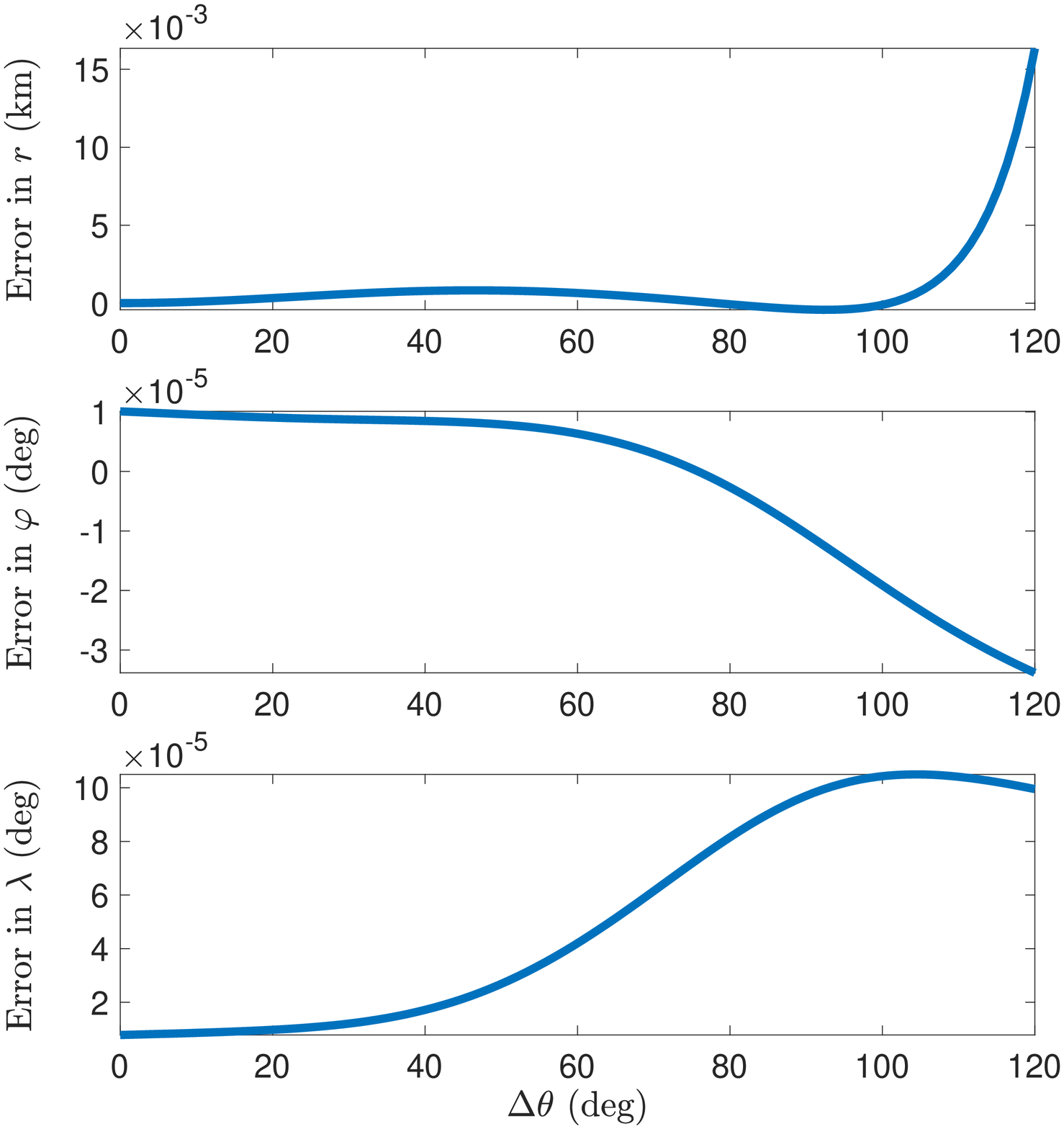}}
	\caption{7th order solution (left) and error (right) in $\{r,\varphi,\lambda\}$ for a hyperbolic orbit.}
	\label{fig:hyp7}
\end{figure}

\begin{figure}[!ht]
	\centering
	{\includegraphics[width = 0.48\textwidth]{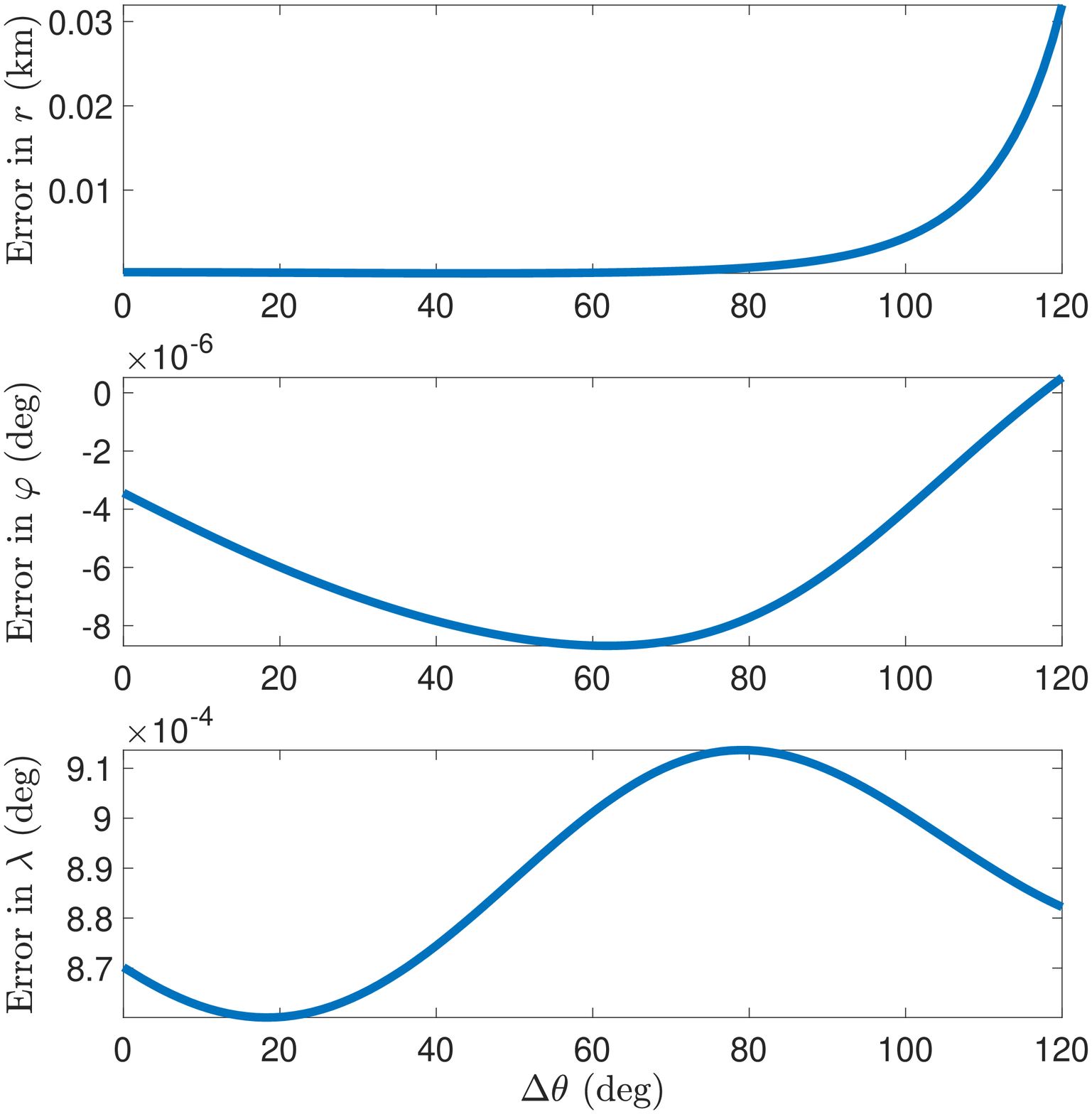}}
	\hspace{0.1cm}
	{\includegraphics[width = 0.48\textwidth]{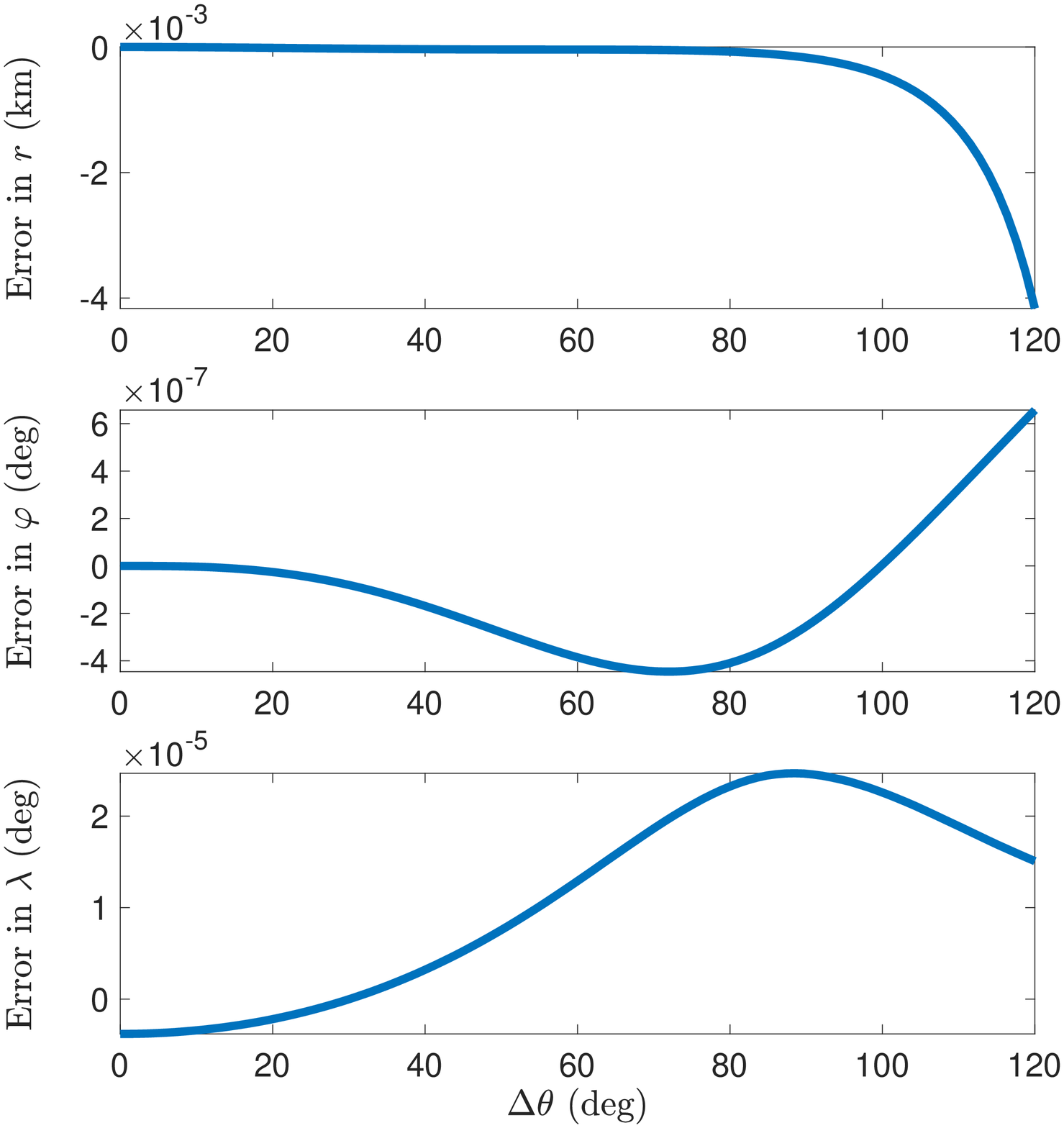}}
	\caption{9th (left) and 11th (right) order solution error in $\{r,\varphi,\lambda\}$ for a hyperbolic orbit.}
	\label{fig:hyp911}
\end{figure}

For the example of hyperbolic orbit, we select the following initial conditions: $a = -35000.0$ km, $e = 1.2$, $inc = 50.0 \deg$, $\omega = 0.0 \deg$, $\Omega = 0.0 \deg$ and $\nu = 0.0 \deg$. Figure~\ref{fig:hyp7} shows the results of the propagation and associated errors for a Koopman solution using basis functions of order 7, while Figure~\ref{fig:hyp911} does the same for orders 9 (left) and 11 (right). Note that in this case, the propagation has been extended just to $\theta = 120^{\circ}$ in order to prevent the variables to become extremely large. Nevertheless, it can be seen from the figures that the precision of the solution is also maintained in this case, having a maximum error on the order of magnitude of meters even for the hyperbolic case.

Note that sometimes applying a set of higher order basis functions produces a slightly decrease in accuracy in the methodology, especially in the longitude of the orbit. This can be seen, for instance, in the errors in latitude of Figure~\ref{fig:circ911} for the cases of 9th and 11th orders, or in error in latitude for this hyperbolic example for orders 7 and 9 (Figures~\ref{fig:hyp7} and~\ref{fig:hyp911}). However, if the order of the basis functions is further increased, this anomalous behavior disappears. An example of that can be observed in Figure~\ref{fig:hyp911} in the latitude of the solution for basis functions of orders 9 and 11. This happens due to numerical errors appearing during the eigendecomposition of the Koopman matrix and subsequent inversion of the matrix of eigenvectors.

\newpage
\subsection{Near-equatorial orbit}

\begin{figure}[!ht]
	\centering
	{\includegraphics[width = 0.48\textwidth]{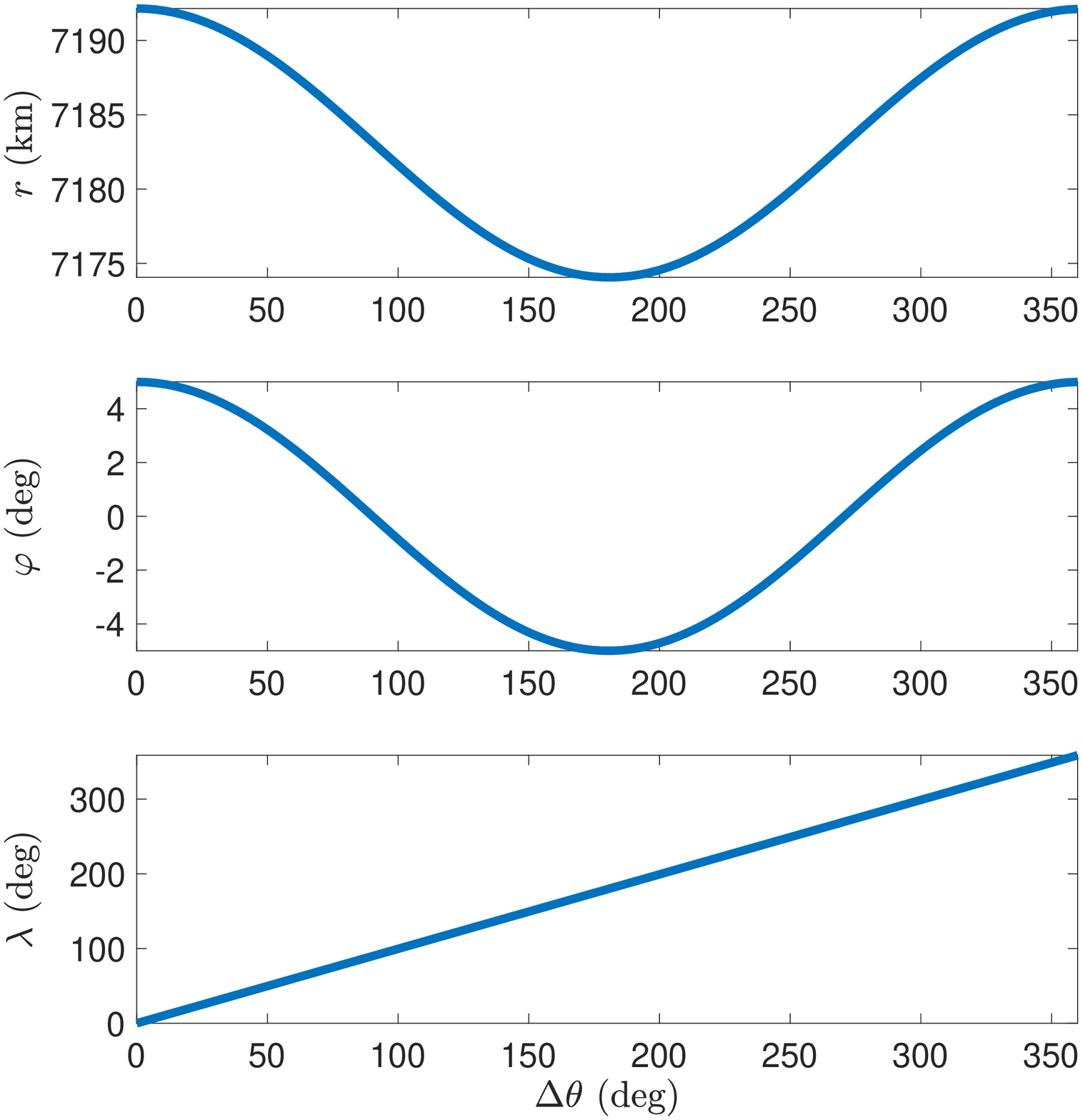}}
	\hspace{0.1cm}
	{\includegraphics[width = 0.48\textwidth]{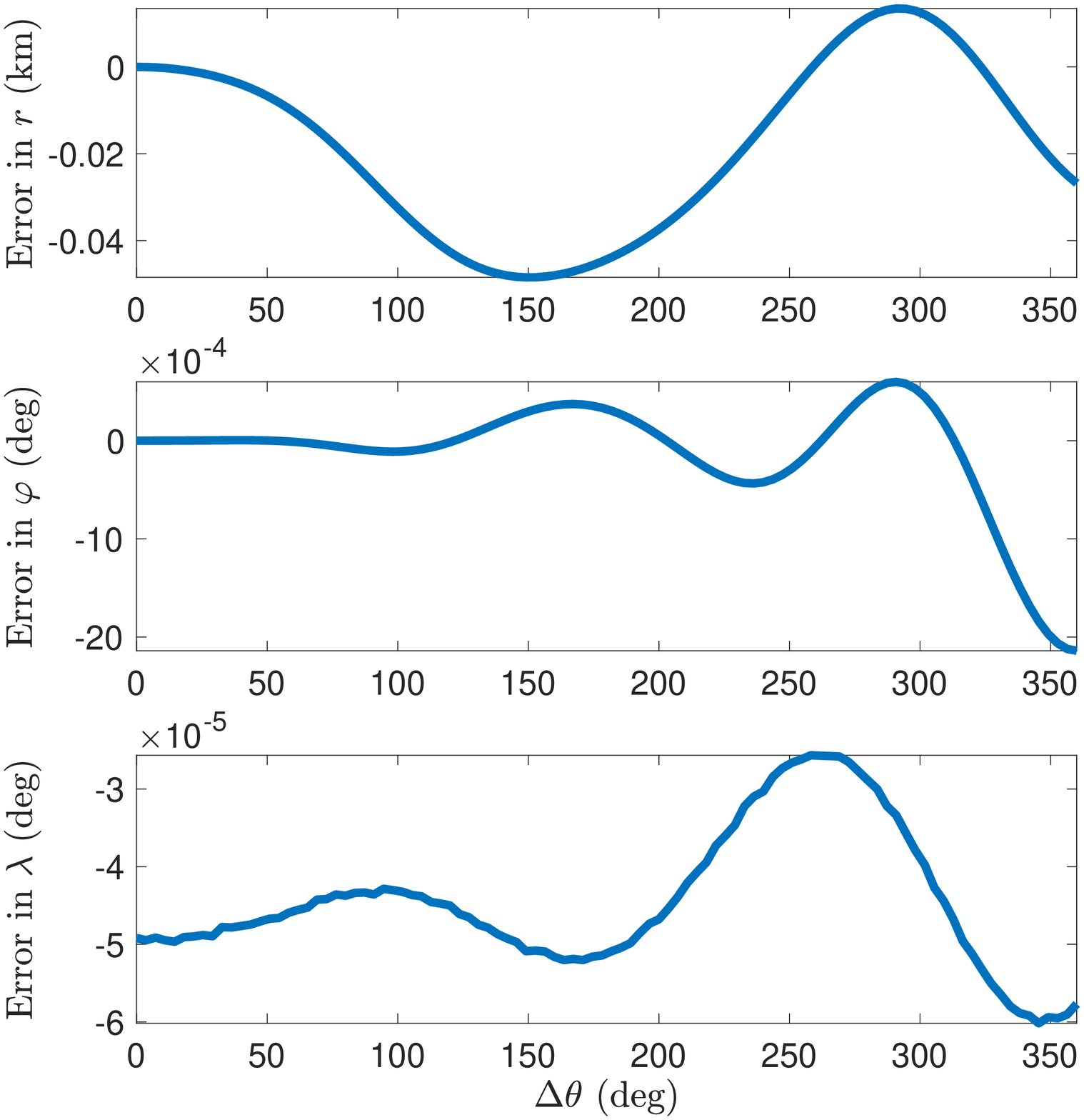}}
	\caption{7th order solution (left) and error (right) in $\{r,\varphi,\lambda\}$ for a $5 \deg$ inclination orbit.}
	\label{fig:eq7}
\end{figure}

\begin{figure}[!ht]
	\centering
	{\includegraphics[width = 0.48\textwidth]{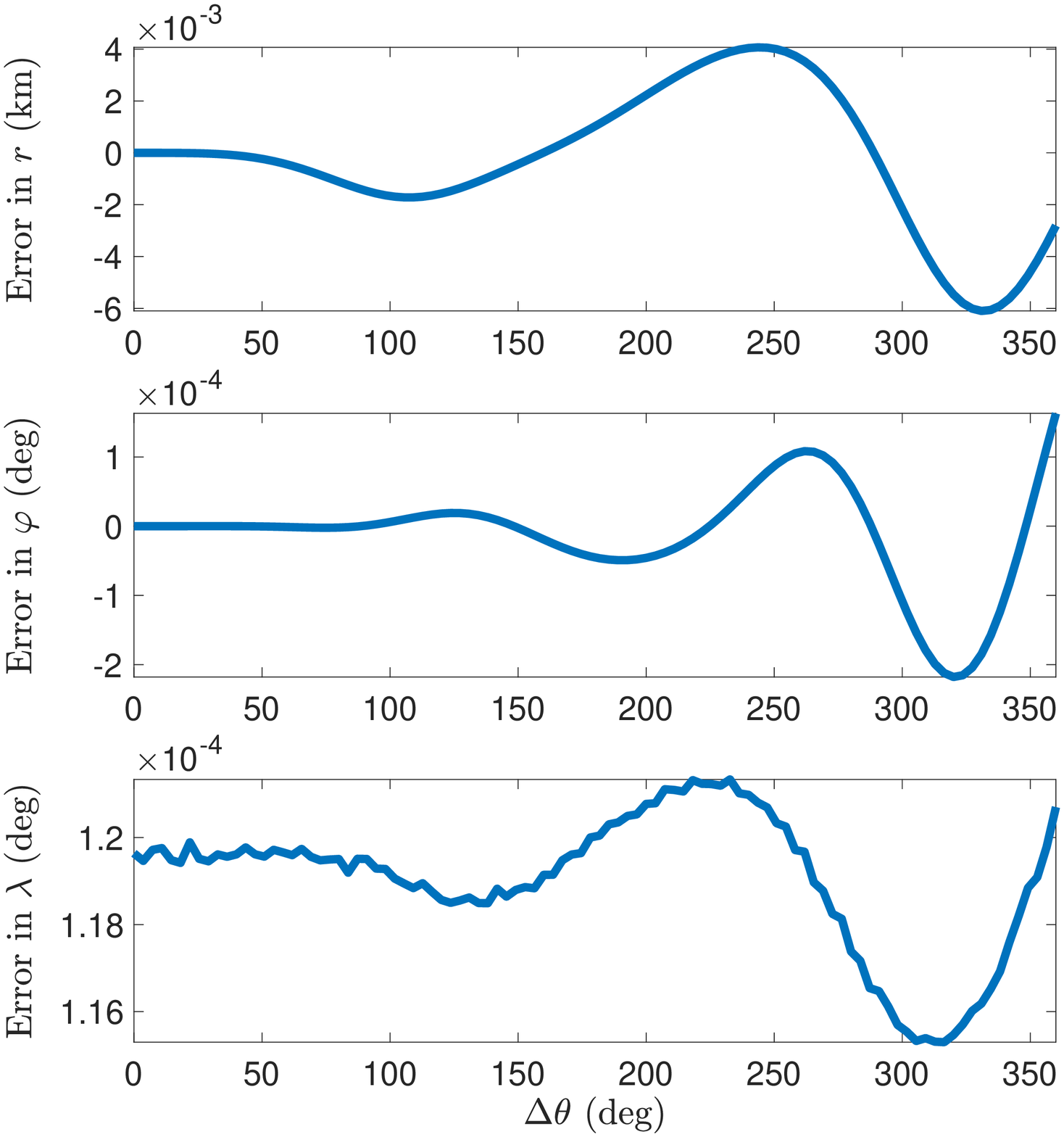}}
	\hspace{0.1cm}
	{\includegraphics[width = 0.48\textwidth]{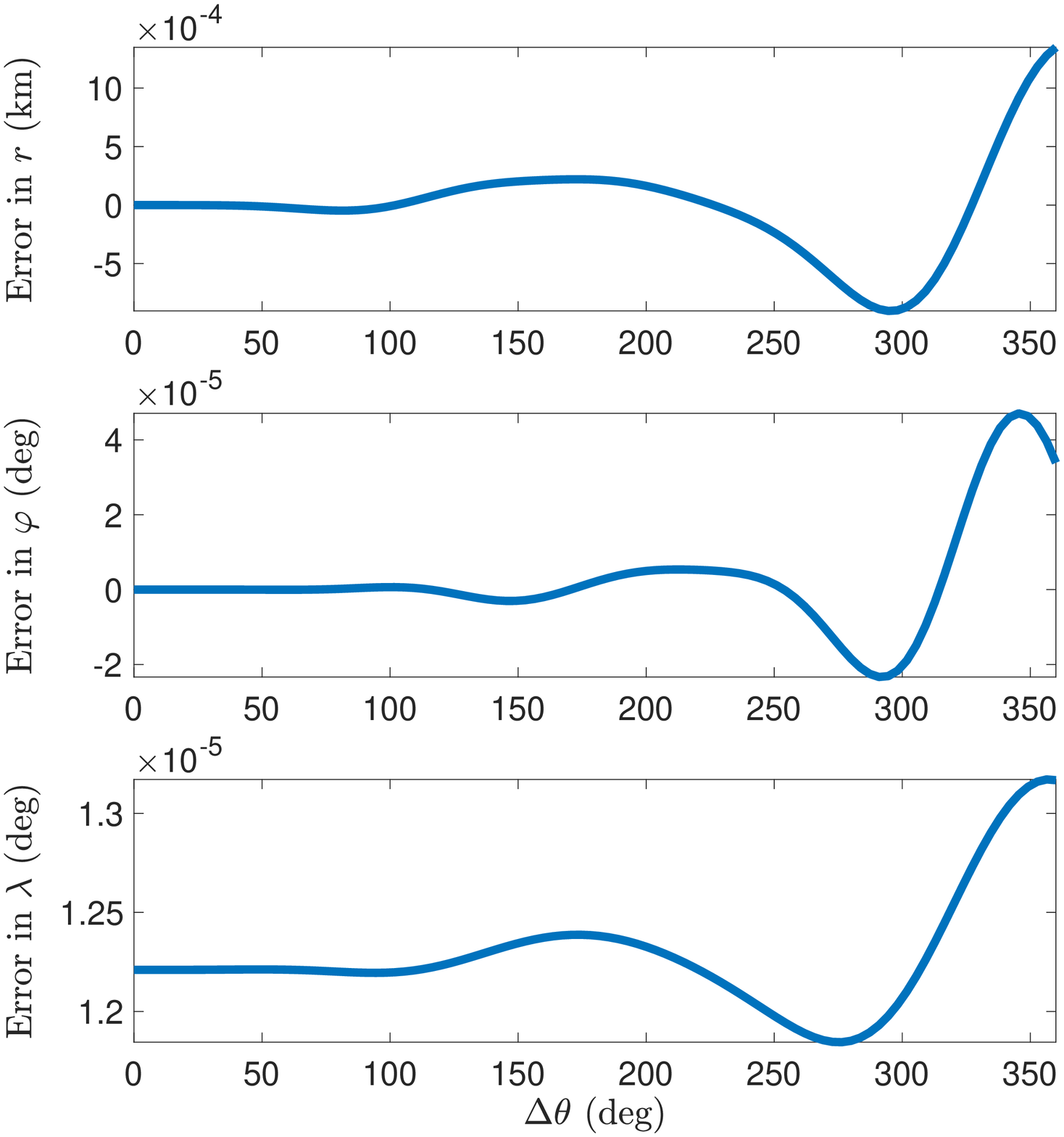}}
	\caption{9th (left) and 11th (right) order solution error in $\{r,\varphi,\lambda\}$ for a $5 \deg$ inclination orbit.}
	\label{fig:eq911}
\end{figure}

This final example is included to show the performance of the formulation for close-to-equatorial orbits. Particularly, the initial conditions selected for this examples are: $a = 7192.15$ km, $e = 0$, $inc = 5 \deg$, $\omega = -90 \deg$, $\Omega = 0.0 \deg$ and $\nu = 180.0 \deg$. As in the previous examples, Figures~\ref{fig:eq7} and~\ref{fig:eq911} show the evolution of the position and the errors for the Koopman operator methodology when using basis functions of orders 7, 9 and 11 respectively. These figures show that the errors are of the same order of magnitude as the ones seen in the previous examples, which presents the performance of this alternative formulation when applied to near-equatorial orbits.

\begin{figure}[!ht]
	\centering
	{\includegraphics[width = 0.48\textwidth]{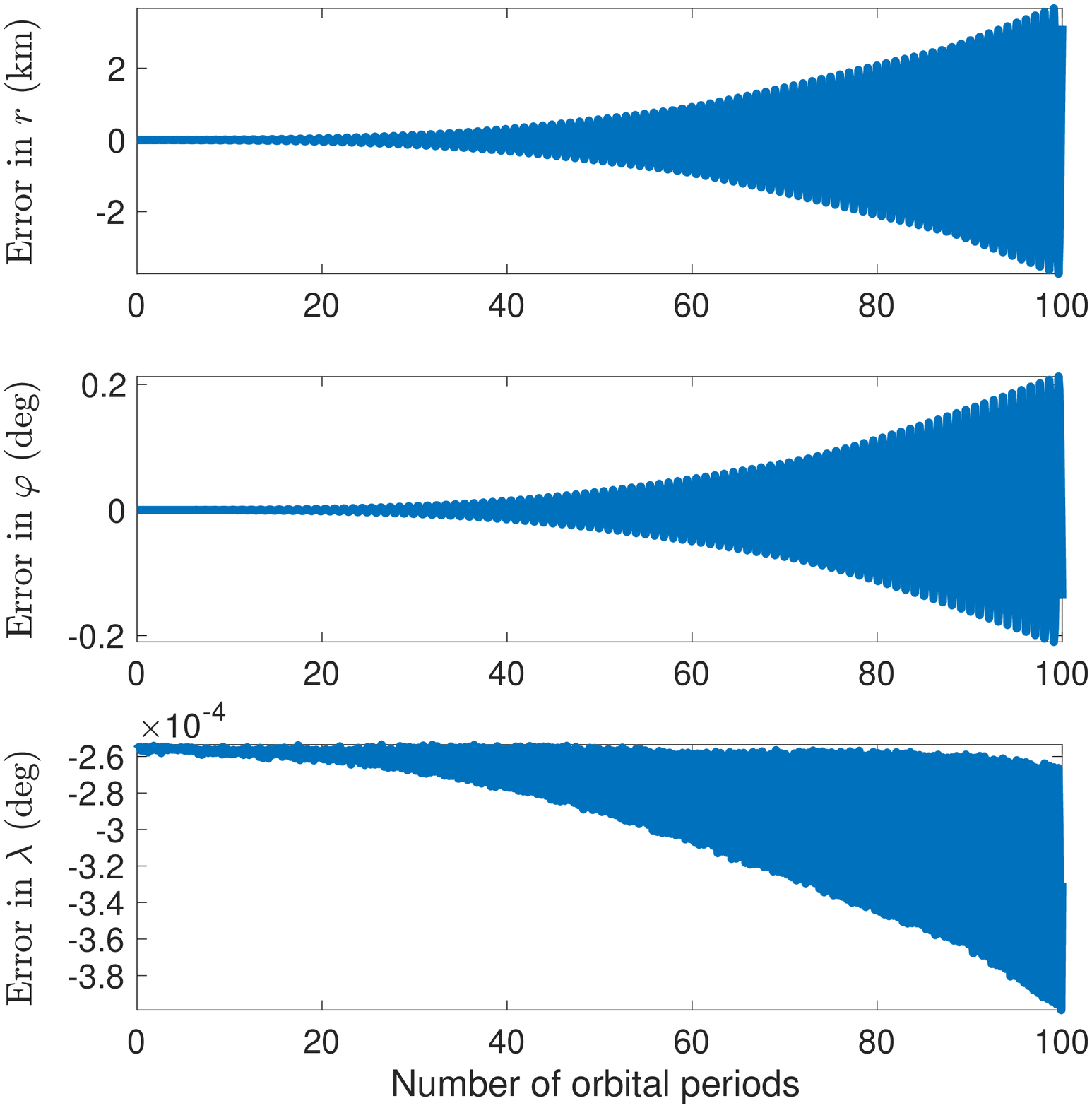}}
	\hspace{0.1cm}
	{\includegraphics[width = 0.48\textwidth]{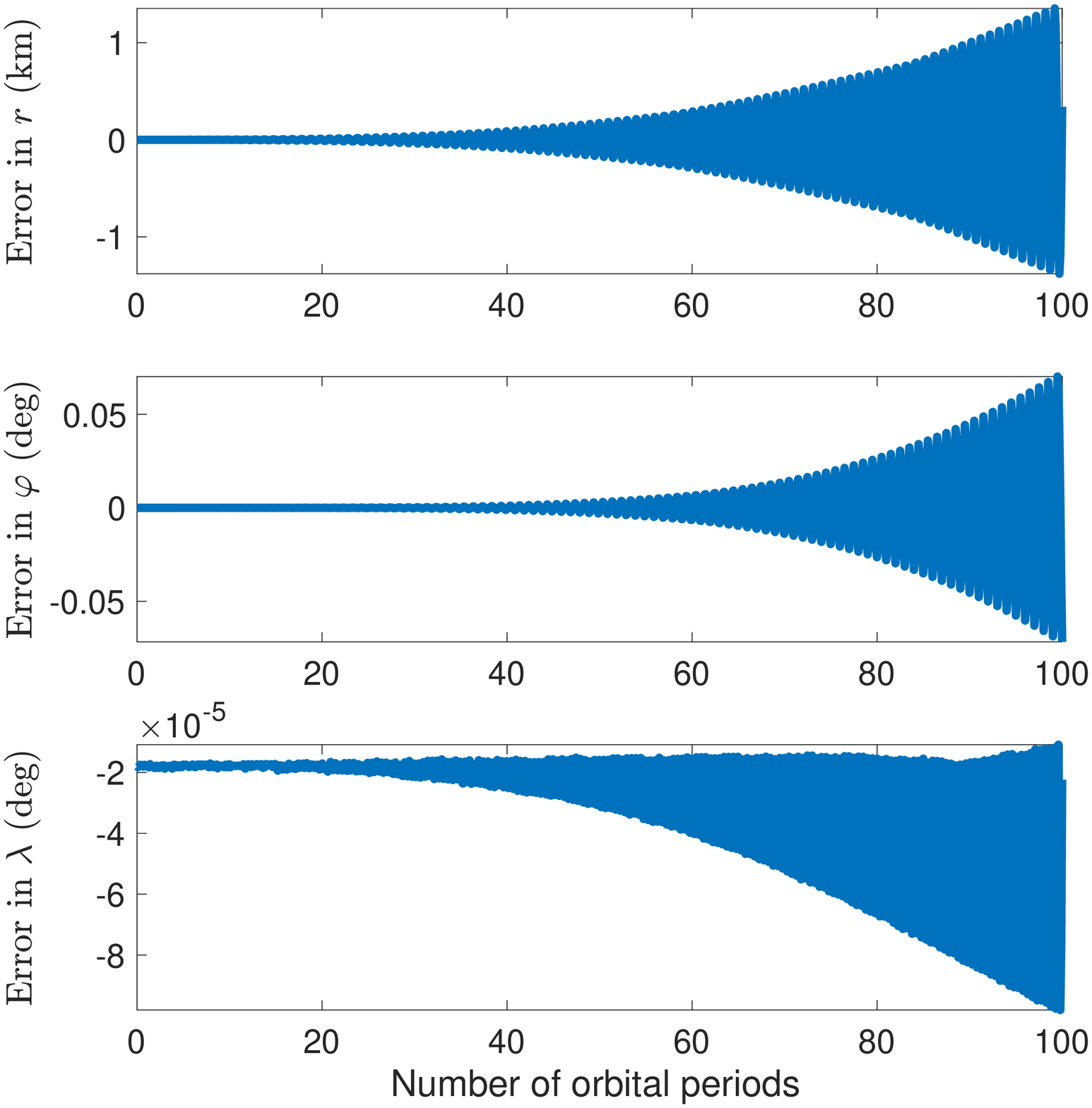}}
	\caption{7th (left) and 9th (right) order solution error for a long term propagation in $\{r,\varphi,\lambda\}$ for the near-equatorial orbit.}
	\label{fig:eqlong}
\end{figure} 

\begin{figure}[!ht]
	\centering
	{\includegraphics[width = 0.48\textwidth]{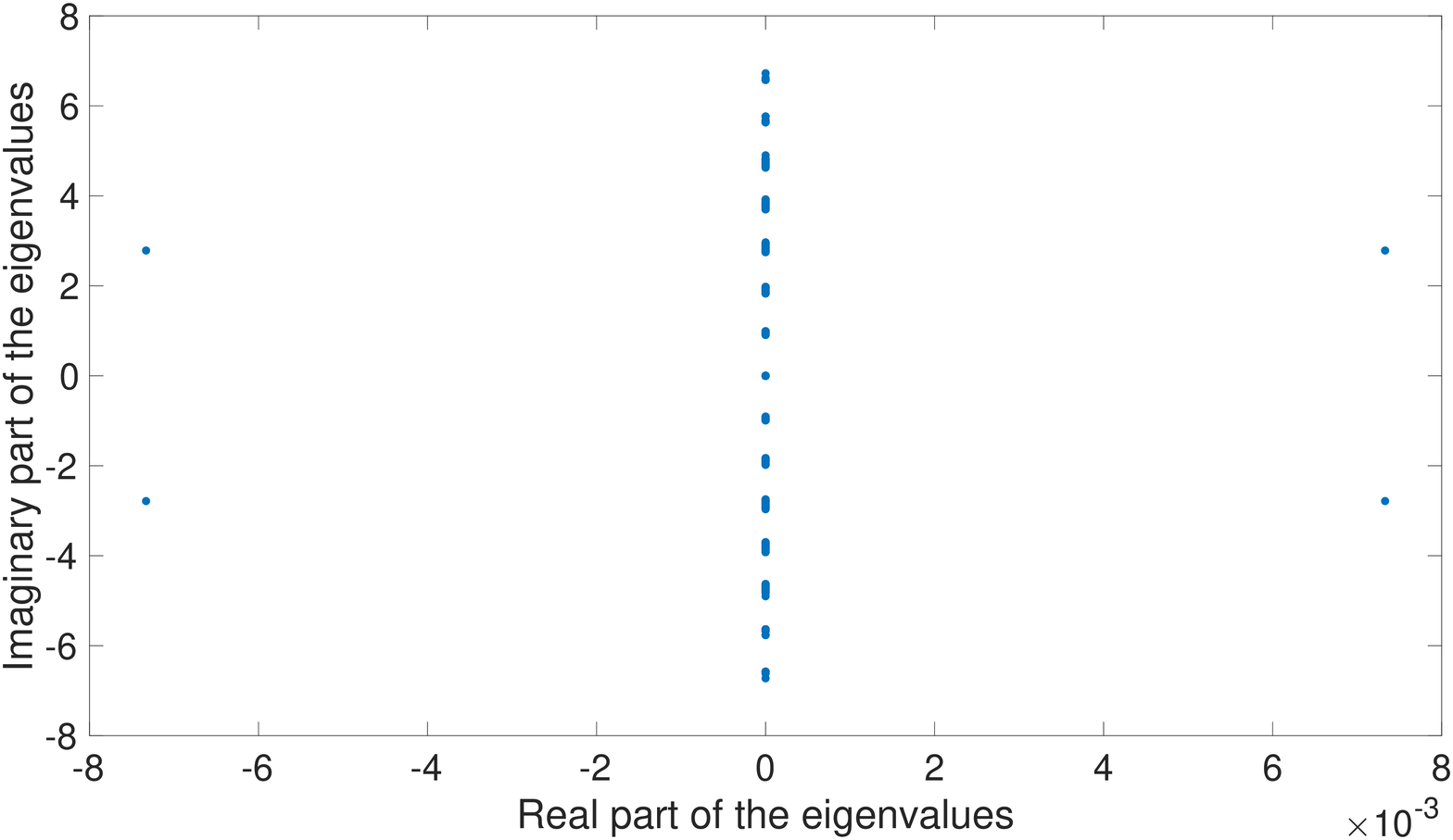}}
	\hspace{0.1cm}
	{\includegraphics[width = 0.48\textwidth]{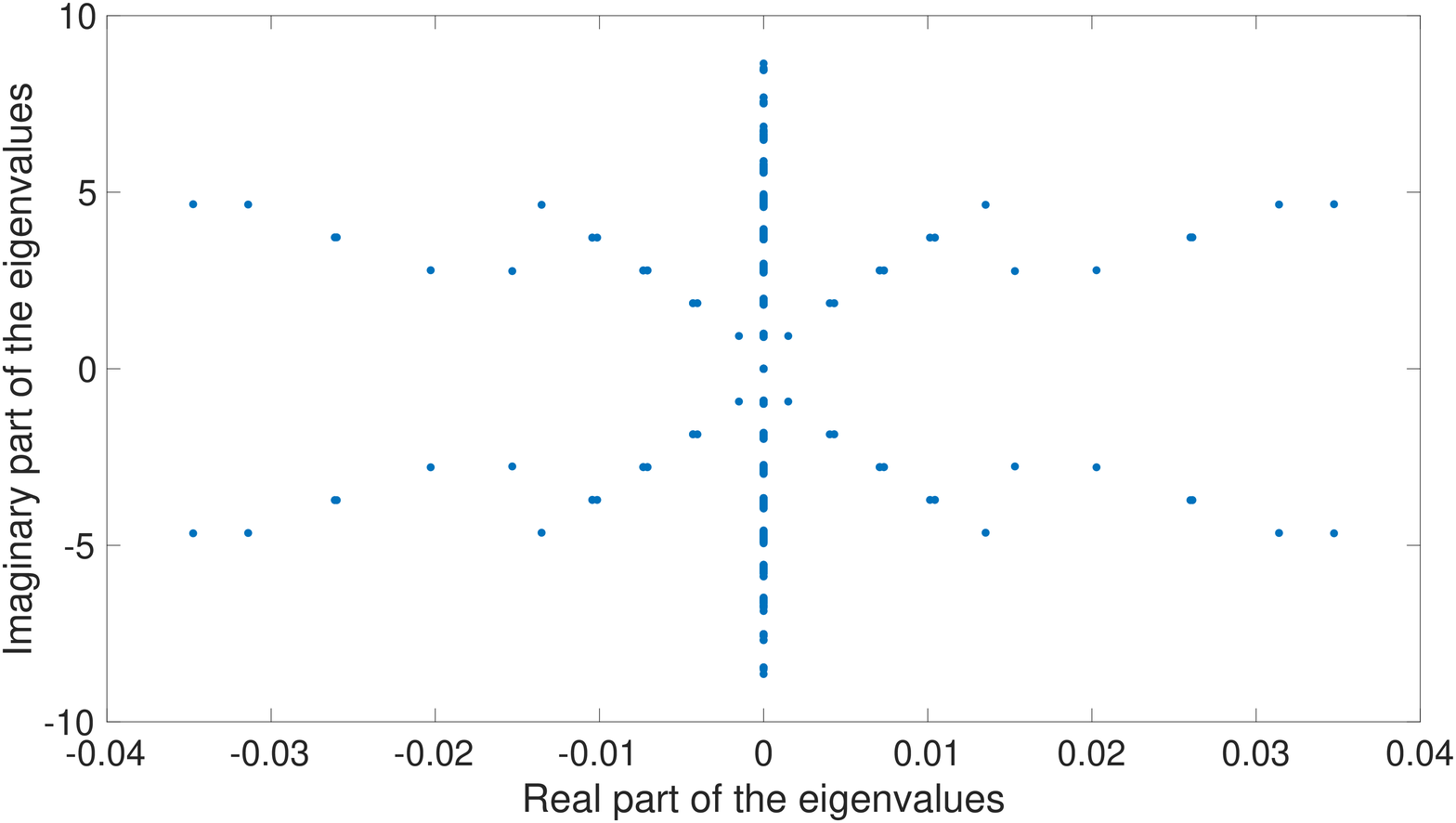}}
	\caption{7th (left) and 9th (right) order eigenvalues for the close-to-equatorial formulation.}
	\label{fig:eigen_eq}
\end{figure}

Additionally, Figure~\ref{fig:eqlong} shows the error of the Koopman operator for 100 orbital revolutions when using 7th order (left), and 9th order (right) basis functions for the subspace representation of the system. The solution presents a maximum error in position of $6$ km (when considering the combined effect of radial distance and angles) after 100 orbital revolutions. Compared to the other cases of study, we can observe a better accuracy performance in the long term. This effect is mainly generated by the lower number of eigenvalues with positive real part than in the general case formulation (see also Figure~\ref{fig:eigen_eq}), making the long term behavior of the solution numerically more stable. Moreover, it is important to note that the accuracy presented in this example, as well as in the other cases of study, can be further improved by increasing the order of basis functions used. Another possibility is to perform a variable normalization based on the estimated variation of the states.

Finally, and since the Koopman matrix is different from the one used in the previous examples (because we are using a slightly different formulation), it is interesting to study the difference in the distribution of eigenvalues of this system. In that regard, Figure~\ref{fig:eigen_eq} shows a map containing all the eigenvalues when using eigenfunctions of order 7 and 9 respectively. As can be seen, the distribution is different from Figure~\ref{fig:eigen}, particularly the real parts of the eigenvalues. This is due to the different nature of both formulations. In the general case, we have as an orbital element a variable whose secular value matches the one of the right ascension of the ascending node. However, in the close-to-equatorial case, that variable was substituted by another whose derivative was linear with the ratio $p_{\lambda}/p_{\theta}$. Additionally, the close-to-equatorial formulation does not require the additional variable $\chi$ which also generates a large number of these eigenvalues with non-zero real part. This shows the importance of selecting an adequate formulation for the use alongside with the Koopman operator since it greatly depends on the eigendecomposition of the system.   

\section{Conclusion}

This paper focuses on the application of the Koopman operator theory to the zonal harmonics problem of a satellite orbiting an oblate celestial body. This represents the first attempt of using operator theory in this important problem in astrodynamics. To that end, this work presents a summary of the Koopman operator theory and how to apply it effectively into the zonal harmonics problem using a set of modified orbital elements that allow to represent any kind of orbit, including, circular, elliptic, parabolic and hyperbolic orbits.

The Koopman operator methodology allows us to obtain in an automated process any order of the solution by an extension of the set of basis functions used to represent the solution. Therefore, this methodology allows the error to be adjusted to any precision required. In that regard, four examples of application are shown in this paper, showing the performance results of this methodology under the effect of the $J_2$ perturbation to very characteristic orbits in celestial mechanics, including a sun-synchronous frozen orbit, a Molniya orbit, a hyperbolic orbit and a close-to-equatorial orbit. These examples show that it is possible to obtain relative errors of $10^{-5}$ when using this methodology for the different cases of application. Similar performance have been observed for other orbits in the same range of altitudes. In that regard, it is important to note that as the altitude of the orbit increases, the intensity of the perturbation produced by the zonal harmonic terms of the gravitational potential is also reduced, and thus, the accuracy performance of the methodology also improves accordingly. For instance, for a geo-synchronous orbit, and when considering only the $J_2$ perturbation, less than 1 cm errors were observed for the same order of basis functions considered in this work.

Compared to other methodologies that focus on the zonal harmonics problem, the Koopman operator directly provides the osculating variation of all the variables involved in the problem, as opposed to other approaches where it is required to define some intermediaries or the transformation of orbital elements. Additionally, the Koopman operator allows us to handle automatically any order of the zonal harmonics of the gravitational potential, not requiring to modify the methodology in any way to obtain the solution. Another interesting result of the Koopman operator is that the methodology provides the spectral behavior of the system, which can be used to study some properties of the problem. An example of that is the searching of periodic orbits, which can be used to define frozen orbits under the zonal harmonics perturbation. Finally, and since the Koopman matrix represents the best linear approximation of the problem with the set of basis functions and variable ranges defined, it can be used for a wider range of applications, In particular, control schemes can benefit from the linearity of the dynamical model defined by the Koopman matrix, and the accuracy of the solution provided by the Koopman operator.


\section*{Acknowledgements}

The authors wish to acknowledge the support of this work by the Air Force’s Office of Scientific Research under Contract Number FA9550-18-1-0115. The authors also wish to thank Daniel Jang at Massachusetts Institute of Technology for his help in reviewing and improving the use of English in this paper.

\printbibliography

\end{document}